\documentclass{article}

\usepackage{arxiv}

\usepackage[utf8]{inputenc} 
\usepackage[T1]{fontenc}    
\usepackage{hyperref}       
\usepackage{xurl}            
\usepackage{booktabs}       
\usepackage{amsfonts}       
\usepackage{nicefrac}       
\usepackage{microtype}      
\usepackage{lipsum}		
\usepackage{graphicx}
\usepackage{doi}

\usepackage[T1]{fontenc}
\usepackage[utf8]{inputenc}
\usepackage{calc}
\usepackage{indentfirst}
\usepackage{fancyhdr}
\usepackage{graphicx,epstopdf}
\usepackage{lastpage}
\usepackage{ifthen}
\usepackage[right]{lineno}
\usepackage{float}
\usepackage{amsmath}
\usepackage{amssymb} 
\usepackage{setspace}
\usepackage{enumitem}
\usepackage{mathpazo}
\usepackage{booktabs} 
\usepackage{titlesec}
\usepackage{xcolor, colortbl} 
\usepackage{multirow}

\usepackage{xspace}
\newcommand{\ifb}{\ensuremath{{\rm fb}^{-1}}\xspace}
\newcommand{\mtop}{\ensuremath{m_{\mathrm{top}}}\xspace}

\newcommand{\ttbar}{\ensuremath{t\bar{t}}\xspace}
\newcommand{\gev}{\ensuremath{\rm GeV}\xspace}
\newcommand{\tev}{\ensuremath{\rm TeV}\xspace}
\newcommand{\ptmiss}{\ensuremath{p_{\mathrm T}^{\mathrm miss}}\xspace}
\newcommand{\met}{\ensuremath{p_{\mathrm{T}}^{\mathrm{miss}}}\xspace}
\newcommand{\sqrts}{\ensuremath{\sqrt{s}}\xspace}

\usepackage{setspace}
 \usepackage{caption}
\title{Dark Matter Searches with Top Quarks}

\date{} 					

\author{ \href{https://orcid.org/0000-0002-5501-4640}{\includegraphics[scale=0.06]{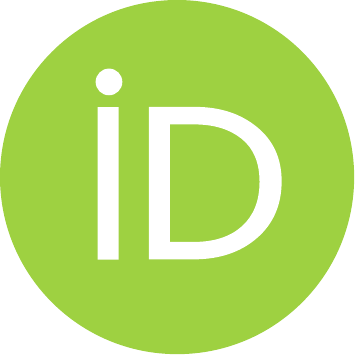}\hspace{1mm}J. Katharina Behr}
    \\
	Deutsches Elektronen-Synchrotron DESY\\
	Notkestr. 85\\
	22607 Hamburg \\
	\texttt{katharina.behr@desy.de} \\
	\And
	\href{https://orcid.org/0000-0003-0748-8494}{\includegraphics[scale=0.06]{orcid.pdf}\hspace{1mm}Alexander Grohsjean} \\
	Universit\"at Hamburg \\
	Luruper Chaussee 149\\
	22761 Hamburg \\
	\texttt{alexander.grohsjean@desy.de} \\
 }

\renewcommand{\undertitle}{Review article}

\begin{document}
\maketitle

\begin{abstract}
Collider signatures with top quarks provide sensitive probes of dark matter (DM) production at the Large Hadron Collider (LHC). In this article, we review the results of DM searches in final states with top quarks conducted by the ATLAS and CMS Collaborations at the LHC, including the most recent results on the full LHC Run 2 dataset. We highlight the complementarity of DM searches in final states with top quarks with searches in other final states in the framework of various simplified DM models. A re-interpretation of a DM search with top quarks in the context of an effective-field theory description of scalar dark energy is also discussed. Finally, we give an outlook on the potential of DM searches with top quarks in LHC Run 3, at the high-luminosity LHC, and possible future colliders. In this context, we highlight new benchmark models that could be probed by existing and future searches as well as those that predict still uncovered signatures of anomalous top-quark production and decays at the LHC.
\end{abstract}


\vspace{1.5cm}

\section{Introduction}
The particle nature of dark matter (DM) is one of the major puzzles in modern particle physics, despite long-standing evidence for its existence.
As early as 1884, Lord Kelvin realised that the mass of the  Milky Way derived from the velocity dispersion of the stars orbiting its centre is very different from the mass of the visible stars. He considered the majority of stars in our galaxy to be dark bodies. 140 years later overwhelming astronomical and cosmological evidence has been accumulated for the existence of Dark Matter (DM) across different scales, ranging from the rotational velocity of stars in ultra-faint galaxies over gravitational lensing effects to precision measurements of the cosmic microwave background~\cite{lensingone,lensingtwo,lensingthree, cmbone, cmbtwo}.

It is well established that 85\% of the matter in our Universe consists of DM. The dominant part of DM must be stable with a lifetime much longer than the age of the Universe. The fact that DM was already produced in the early Universe may provide a clue to non-gravitational interactions. At the same time, the feature that DM must form cosmological structures consistent with current observations allows setting a limit on the strength of DM interactions with SM particles and with itself. It is clear that none of the Standard Model particles is consistent with all of these observations.

One of the highly-motivated theory paradigms for DM is the so-called WIMP (weakly interacting massive particle) paradigm, also known as the WIMP miracle~\cite{wimp}. Assuming DM to be produced via the freeze-out mechanism, one can achieve the observed relic density when the DM mass is close to the electroweak scale and when the DM coupling to Standard Model particles is at the order of the weak interaction. Consequently, DM particles could be produced and studied at the Large Hadron Collider (LHC)~\cite{LHC}.

Besides producing DM under controlled experimental conditions, the LHC would also provide access to the particles mediating the interactions between DM and the Standard Model. A DM mediator produced in proton-proton ($pp$) collisions could decay to DM particles. Such \textit{invisible decays} could only be inferred via the presence of missing transverse momentum, \met, in the detector. However, a DM mediator decaying back into SM particles (\textit{visible decays}) would provide direct access to its properties. DM searches at the LHC explore both avenues. To detect the invisible decays of a mediator, it is mandatory to produce the mediator in association with SM particles. In this review article, we will focus on the associated production with top quarks and more generally on the role of top quarks in the quest for DM. Best suited to study DM in top quark channels are the two general purpose detectors ATLAS~\cite{atlas} and CMS~\cite{cms}.  

Discovered in 1995 at the Fermilab Tevatron collider~\cite{topcdf,topd0}, the top quark is the heaviest of all known elementary particles. In the case of a DM mediator with Yukawa-like couplings, the top quark would be ideal for discovery. Moreover, the top quark would allow for a first characterization of the dark sector. Due to its short lifetime, the top quark fully transmits its spin information to the decay particles. In turn, this allows inferring the spin of the mediator for both the associated production of top quarks and DM as well as for the decay of a mediator to a top-quark pair.

Another major unknown in the physics of our universe, beside the particle nature of DM, is the origin of its accelerating expansion~\cite{SupernovaSearchTeam:1998fmf,SupernovaCosmologyProject:1998vns}, which is usually attributed to the presence of a yet unknown repulsive force, referred to as dark energy (DE). If DE is a scalar field, it may be possible to produce it at the LHC. Like DM, DE would escape the detector unnoticed. DM searches with top quarks could be sensitive to DE production, as shown in Sections~\ref{sec:DE_model} and~\ref{sec:DE_results} of this review.

This article is structured as follows. After a detailed discussion of the underlying DM models in Section~\ref{sec:models}, we will focus on the experimental signatures of DM searches involving top quarks at LHC in Section~\ref{sec:signatures}. In Section~\ref{sec:results}, current highlights and results from DM searches at LHC are summarised. We conclude with a discussion of uncovered signatures and models, follwoed by an outlook on prospects for discovering DM at future colliders in Sections~\ref{sec:discussion} and \ref{sec:outlook}. 
\section{Models with BSM signatures involving top quarks}\label{sec:models}
Collider searches for DM are usually interpreted in the context of so-called \textit{simplified models}, which contain a minimal set of new particles and couplings. Most of these models contain only a single Dirac DM particle and a single mediator particle. They are characterised by a minimal set of free parameters, namely the masses of the DM and mediator particles and the couplings of the mediator to the SM and dark sector. Simplified models provide a convenient framework to compare searches in different final states and among different experiments.
In the following, the simplified models used for the interpretation of DM searches involving top quarks are described. Additionally, an effective-field theory (EFT) description of scalar DE is introduced.
\subsection{Vector and axial-vector mediators}
\subsubsection{Flavour-conserving interaction}
\label{sec:AVV_model}

A mediator with flavour-universal couplings to the SM quarks and leptons, respectively,
is predicted in a simplified model that describes a flavour-conserving interaction between a fermionic WIMP DM particle $\chi$ and the SM fermions~\cite{Abercrombie:2015wmb}. It is based on a simple extension of the SM by a new $U(1)$ gauge symmetry under which $\chi$ as well as some of the SM fermions are charged, thus allowing the mediator to couple to the SM sector. The interaction described by this gauge group is mediated by the $s$-channel exchange of a new, electrically neutral spin-1 particle $Z'$ with either vector or axial-vector couplings to the DM and SM fields. It will be referred to as \textit{ vector mediator} or \textit{ axial-vector mediator} in the following.

The model contains five free parameters~\cite{Abercrombie:2015wmb}: the masses of the mediator, $m_{Z'}$, and the DM particle, $m_{\chi}$, as well as the quark-flavour universal coupling $g_q$ of the mediator to quarks, the lepton-flavour universal coupling $g_{\ell}$ of the mediator to leptons, and the coupling $g_{\chi}$ of the mediator to DM.

\begin{figure}[h!]
\centering 
\includegraphics[width=0.3\textwidth]{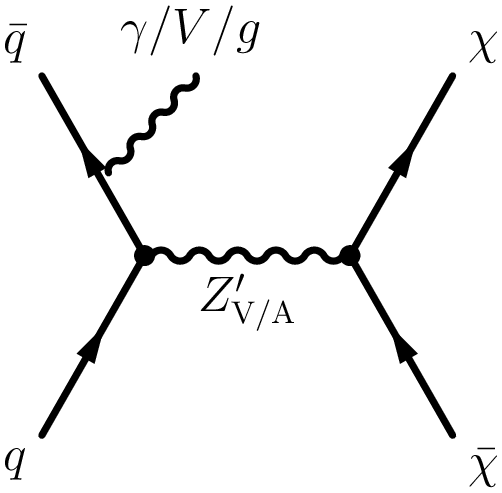}
\includegraphics[width=0.3\textwidth]{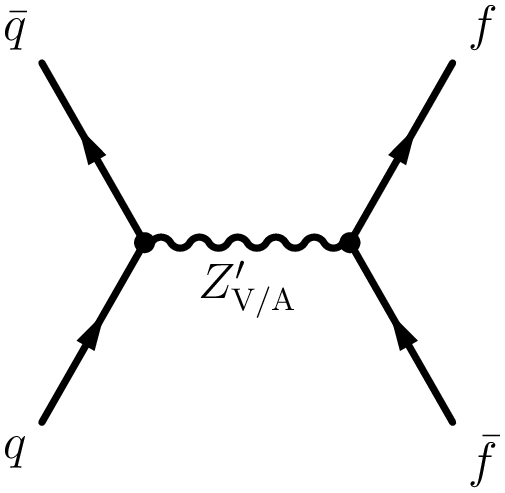}
\caption{Schematic representation of the dominant production and decay modes of the simplified model with an $s$-channel vector or axial-vector mediator $Z'$~\cite{ATLAS:2019wdu}.\label{fig:AVV_Feyn}}
\end{figure}

The mediator can decay either invisibly into a $\chi\bar{\chi}$ pair or visibly into a fermion-anti-fermion $f\bar{f}$ pair, as illustrated schematically by the left and right diagrams, respectively, in Figure~\ref{fig:AVV_Feyn}. The former process can be detected as a \met+$X$ signature in the presence of initial-state radiation (ISR), where $X$ can be a gluon, photon, or vector boson, depending on the type of ISR, while the latter process results in a resonant enhancement in the invariant mass spectrum of the $f\bar{f}$ pair.

Constraints on this model are derived in various parameter planes, including the $(m_{Z'},m_{\chi})$ plane for fixed couplings $g_q$, $g_{\ell}$, $g_{\chi}$~\cite{ATLAS:2019wdu} and as upper limits on $g_q$ as a function of $m_{Z'}$, as shown in Section~\ref{sec:AVV_results}.

\subsubsection{Flavour-changing interaction}
\label{sec:VFC_model}

\begin{figure}[h!]
\centering 
\includegraphics[width=0.3\textwidth]{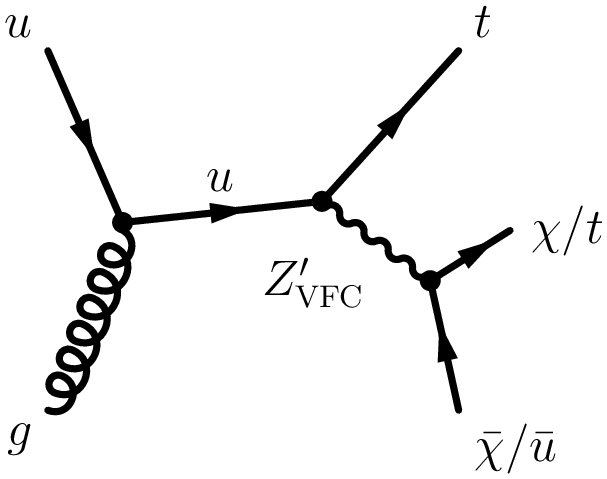}
\includegraphics[width=0.27\textwidth]{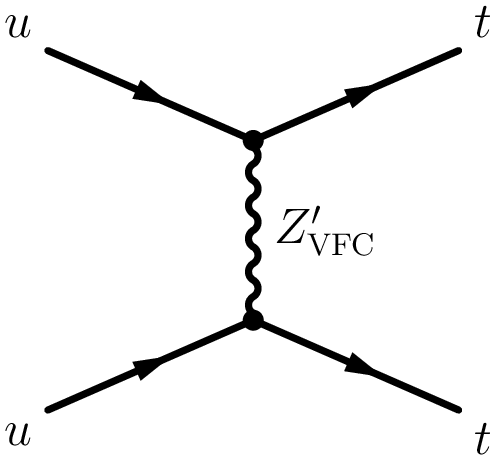}
\includegraphics[width=0.3\textwidth]{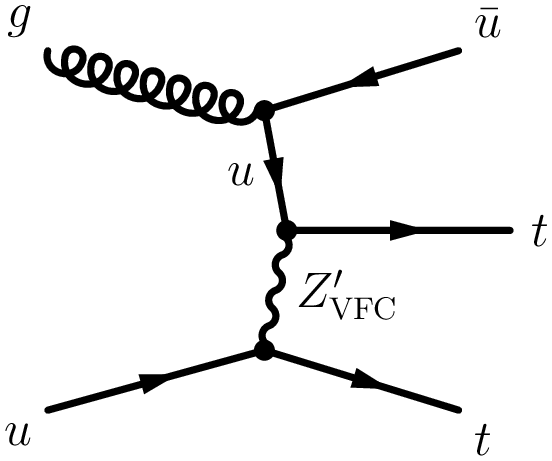}
\caption{Schematic representation of the dominant production and decay modes of the VFC model~\cite{ATLAS:2019wdu}.\label{fig:VFC_Feyn}}
\end{figure}

DM signatures with top quarks are predicted in simplified models containing a vector mediator $Z'_{\mathrm{VFC}}$ with a flavour-changing coupling $V_{ut}$ to the top and up quark.This type of model, referred to as \textit{VFC model} in the following, is motivated, for example, by scenarios with DM in a hidden sector that only interacts with the SM sector via a flavour-changing coupling of a $Z'$ boson~\cite{Boucheneb:2014wza,Kamenik:2011nb}.
The dominant production and decay modes of the VFC model are shown in Figure~\ref{fig:VFC_Feyn}.
The mediator can be produced on-shell in association with a single top or anti-top (left diagram)
and decay either invisibly into DM or visibly into a top and up quark.
The former decay results in a \met+$t$ signature, often referred to as \textit {mono-top}.
The latter decay yields a characteristic final state with two top quarks ($tt$) or two anti-top quarks $\bar{t}\bar{t}$ (same-sign $tt$). This signature can be easily distinguished from the more abundant $t\bar{t}$ production via SM processes by the sign of the lepton charges in fully leptonic decays.
Similar $tt/\bar{t}\bar{t}$ final states arise from the other two diagrams in Figure~\ref{fig:VFC_Feyn},
which represent the $t$-channel exchange of the $Z'_{\mathrm{VFC}}$ mediator.

The VFC model is fully characterised by four free parameters: the mass of the mediator, $m_{Z'_{\mathrm{VFC}}}$, the mass of the DM particle, $m_{\chi}$, the coupling of the mediator to DM, $g_{\chi}$, and the flavour-changing coupling, $g_{ut}$~\cite{ATLAS:2018alq}.
The DM mass has no significant impact on the collider phenomenology of the VFC model, if $2 m_{\chi} < m_{Z'_{\mathrm{VFC}}}$ and is fixed to a value of 1~GeV for existing collider searches~\cite{ATLAS:2019wdu}.
Constraints on the VFC model are accordingly derived in several parameter planes involving the remaining free parameters (or dependent parameters): $m_{Z'_{\mathrm{VFC}}}$, $g_{ut}$, and the invisible branching ratio $\mathcal{BR}(\chi \bar{ \chi})$ of the mediator.

\subsection{Scalar and pseudoscalar mediators}
A preferred coupling of DM to top quarks is predicted in simplified models containing a spin-0 mediator with Yukawa-like couplings to SM fermions. The mediator can be either a scalar ($\phi$) or pseudoscalar ($a$). These models can be straightforwardly embedded in ultra-violet (UV) complete theories with extended Higgs sectors, such as Two-Higgs-Doublet Models (2HDMs, see also Section~\ref{sec:2HDM}). 
Assuming Yukawa-like couplings allows this class of models to satisfy strong constraints from flavour precision measurements. The dynamics of flavour violation are  completely determined by the structure of the ordinary fermion Yukawa couplings, which is referred to as \textit{Minimal Flavour Violation (MFV)}~\cite{DAmbrosio:2002vsn}.

The simplified models described in this section can be broadly categorised into models with a colour-neutral and a colour-charged interaction. An overview of the models falling into each category can be found in Ref.~\cite{ATLAS:2019wdu} and references therein. Two representative benchmark models used by the ATLAS and CMS Collaborations are presented in the following.

\subsubsection{Colour-neutral interaction}
\label{sec:SPS_model}

\begin{figure}[h!]
\centering 
\includegraphics[width=0.24\textwidth]{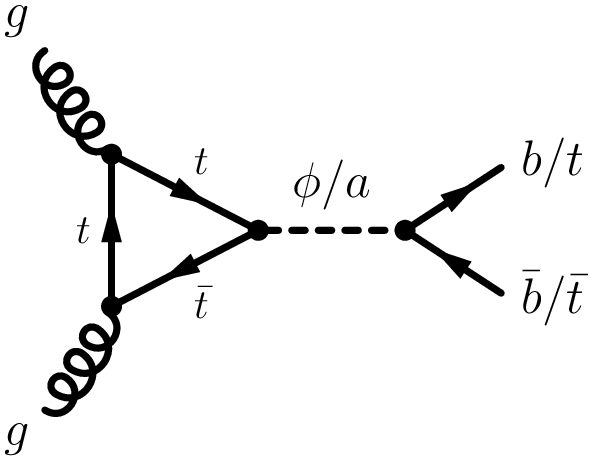}
\includegraphics[width=0.24\textwidth]{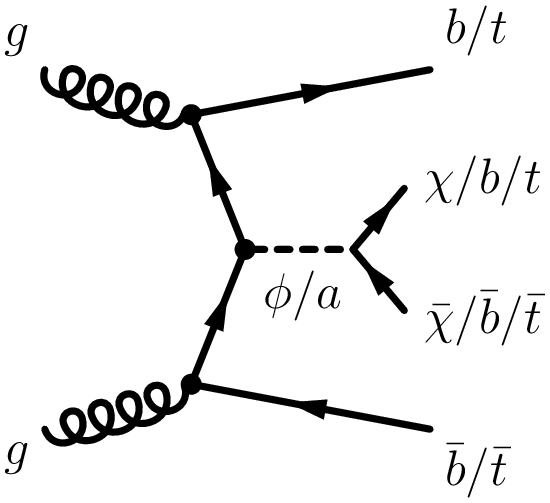}
\includegraphics[width=0.24\textwidth]{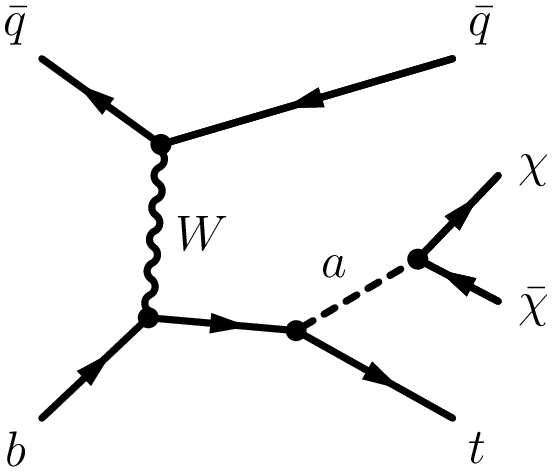}
\includegraphics[width=0.24\textwidth]{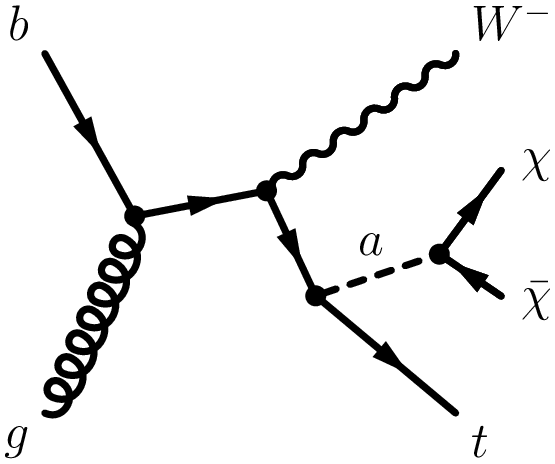}
\caption{Schematic representation of the dominant production and decay modes with heavy-flavour quarks in the final state in the simplified model with a scalar ($\phi$) or pseudoscalar ($a$) mediator~\cite{ATLAS:2019wdu}. 
\label{fig:SPS_Feyn}}
\end{figure}

A colour-neutral interaction between a SM and a DM particle is described by a simplified model with a neutral, scalar or pseudoscalar mediator~\cite{Buckley:2014fba,Abercrombie:2015wmb} with Yukawa-like couplings to the SM fermions. The model has four free parameters: the mass of the DM particle, $m_{\chi}$,
the mass of the mediator, $m_{\phi/a}$, the coupling of the mediator to DM, $g_{\chi}$, and the coupling of the mediator to SM fermions. The latter is parameterised by a flavour-universal coupling constant $g_{q}\equiv g_{u} = g_{d} = g_{\ell}$, which modifies the SM-like Yukawa coupling of the mediator to fermions~\cite{Buckley:2014fba}, thus satisfying the requirements of MFV. It should be noted that couplings to leptons are explicitly included in the model but in practice the related signatures play no significant role in the parameter space accessible to collider searches~\cite{Abercrombie:2015wmb}. Couplings to vector bosons $W, Z$ are not included in this simplified model~\cite{Buckley:2014fba}.
The Yukawa-like couplings imply that the mediator is mostly produced via loop-induced gluon fusion via a heavy-quark dominated loop or in association with heavy-flavour quarks, mostly top quarks. Additionally, visible decays of the mediator preferentially result in heavy quarks.
The dominant production and decay modes of the mediator with heavy-flavour quarks in the final state are shown in Figure~\ref{fig:SPS_Feyn}. These are (from left to right):
\begin{itemize}
    \item visible decay of a mediator produced via gluon-fusion to heavy-flavour quarks, resulting in a resonant $t\bar{t}$ or $b\bar{b}$ signal;
    \item associated production of a mediator that decays either visibly or invisibly with heavy-flavour quarks, leading to a \met+$t\bar{t}$/$b\bar{b}$ signature in the case of invisible mediator decay or characteristic fully visible $t\bar{t}t\bar{t}$, $t\bar{t}b\bar{b}$, $b\bar{b}b\bar{b}$ signatures;
    \item associated production of an invisibly decaying mediator with a top quark and a light ($d,u,s,c$) quark, leading to a \met+$tj$ signature;
    \item associated production of an invisibly decaying mediator with a top quark and a $W$ boson, resulting in a \met+$tW$ signature.
\end{itemize}

Additional signatures not shown here include \met+jet and \met+$V/h$ production.

It should be noted that, while the Yukawa-like coupling structure implies a greater importance of signatures involving top quarks rather than bottom quarks in the final state, signatures involving bottom quarks are still relevant as some UV completions of this simplified model involve a parameter modifying the relative importance of the couplings to up- and down-type quarks. In these UV completions, signatures involving bottom quarks can be more sensitive than signatures involving top quarks if the couplings to up-type quarks are suppressed.

\subsubsection{Colour-charged interaction}
\label{sec:SCC_model}

A colour-charged interaction between the SM quarks and DM is described in a class of simplified models containing a scalar, colour-triplet mediator particle. This type of simplified models is inspired by the Minimal Supersymmetric Standard Model (MSSM)~\cite{MSSM1,MSSM2} with first- and second-generation squarks and neutralino DM~\cite{ATLAS:2019wdu}. The mediator couplings to quarks and DM in the simplified models, however, can differ from those of the MSSM, leading to additional production diagrams.

\begin{figure}[h!]
\centering 
\includegraphics[width=0.285\textwidth]{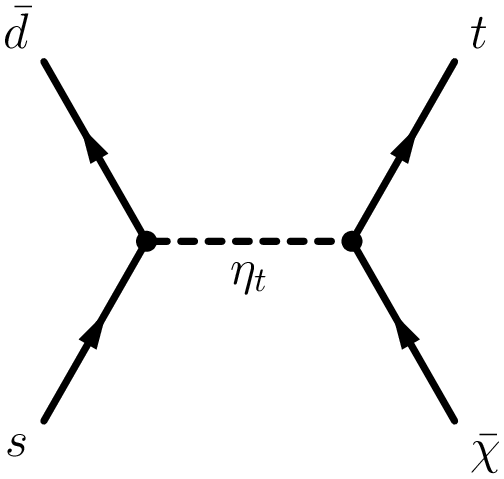}
\caption{Schematic representation of \met+$t$ production via a colour-changing scalar mediator $\eta_t$~\cite{ATLAS:2019wdu}. 
\label{fig:SCC_Feyn}}
\end{figure}

Different models of colour-charged mediators, differing by the mediator couplings to quarks, have been probed at the LHC. These include a model with preferred couplings of the mediator to the first and second quark generation, a model with preferred mediator couplings to bottom quarks, and a model with preferred mediator couplings to top quarks. Only the latter will be discussed in this review.
The concrete realisation of this model is documented in Ref.~\cite{Boucheneb:2014wza}. It contains a new SU(2)$_{\mathrm{L}}$ singlet field that couples to right-handed quarks.
The mediator corresponding to this field is produced from a down-type quark-anti-quark pair and decays to a top quark and a DM particle, as illustrated in Figure~\ref{fig:SCC_Feyn}.
This model can be related to the MSSM if an additional R-parity violating interaction of the
top squark with the down-type quarks is assumed~\cite{ATLAS:2019wdu}.
The free parameters of this model are the mass of the DM particle, $m_{\chi}$, the mass of the mediator, $m_{\eta_t}$, the $t$-DM coupling strength of the mediator, $\lambda_t$, and the coupling strength of the mediator to down-type quarks, $g_{ds}$.

\subsection{Extended Higgs sectors}
\label{sec:2HDM}

Extended Higgs sectors are predicted by a range BSM theories, such as supersymmetry~\cite{Djouadi:2008gy}, certain classes of axion models~\cite{PDG2020}, or theories predicting additional sources of CP violation in the Higgs sector to explain the observed baryon asymmetry in the universe~\cite{Carena:2015uoe,Fuchs:2017wkq}.
Extension of the SM Higgs sector by a second complex SU(2) doublet, referred to as Two-Higgs-Doublet Models (2HDMs), are among the simplest and most studied models with an extended Higgs sector, historically due to their strong motivation from supersymmetry.
In the past years, 2HDMs have also received considerable attention from the DM community as a means of embedding the simplified, mediator-based, models described in the previous sections in the context of a UV-complete and renormalisable framework with a broader collider phenomenology.
Models of DM based on a 2HDM with a vector~\cite{Berlin:2014cfa}, pseudoscalar~\cite{Bauer:2017ota,Goncalves:2016iyg}, and scalar~\cite{Bell:2016ekl} mediator have been proposed. Concrete realisations of the former two have been used as benchmark models by the LHC experiments.
Models with vector mediators are not discussed in this review as final states with top quarks do not play a dominant role in their phenomenology.
Models with a pseudoscalar mediator, on the other hand, feature a rich phenomenology involving relevant signatures with top quarks due to the Yukawa-type coupling of the mediator to SM fermions.
pseudoscalar mediators are also particularly interesting to study at the LHC as they are not strongly constrained by direct-detection experiments because the DM-nucleon scattering cross-section pseudoscalar couplings is strongly suppressed at tree-level by the momentum transfer in the non-relativistic limit~\cite{Abe:2018emu}.
A concrete realisation of a 2HDM with a pseudoscalar mediator that is used as a benchmark model by the LHC experiments is described in Section~\ref{sec:2HDMa_model}.

\subsubsection{2HDM with a pseudoscalar mediator}
\label{sec:2HDMa_model}

A 2HDM with a pseudoscalar mediator $a$~\cite{Bauer:2017ota}, referred to as 2HDM+$a$ in the following, is a more complex simplified model that embeds the phenomenology of the simplified models with a colour-neutral pseudoscalar mediator (Section~\ref{sec:SPS_model}) in more complete model with a second complex SU(2) doublet. The 2HDM in this model has a CP-conserving potential with a softly broken $\mathbb{Z}_2$ symmetry~\cite{Gunion:2002zf}.
Its Higgs sector contains five Higgs bosons: two scalars, $h$ and $H$, a pseudoscalar, $A$, and two charged Higgs bosons $H^{\pm}$.
The alignment limit is assumed, meaning that one of the two scalars of the model is identified with the 125~GeV Higgs boson discovered in 2012. Furthermore, the Yukawa structure of the 2HDM is of type-II~\cite{Branco:2011iw} meaning that couplings of the additional Higgs bosons to top quarks are preferred over those to other fermions at low values of the ratio of the two vacuum expectation values, $\tan\beta$, one of the model parameters with the biggest impact on the collider phenomenology of the model.
The pseudoscalar mediator $a$ mixes with the pseudoscalar $A$ of the 2HDM with mixing angle $\theta$.

The phenomenology of the 2HDM+$a$ is fully defined by 14 free parameters, making it considerably more complex than the simplified models described in the previous sections. These parameters are: the masses $m_h$, $m_H$ and $m_A$ of the neutral Higgs bosons; the masses $m_{H^{\pm}}$ of the charged Higgs bosons; the mass $m_a$ of the mediator; the mass $m_{\chi}$ of the DM particle; the coupling $y_{\chi}$ between DM and the mediator; the three quartic couplings $\lambda_{\textrm{P1}}$, $\lambda_{\textrm{P2}}$, $\lambda_3$ of the mediator to the SU(2) fields; the vacuum expectation value (VEV) $v$ of the electroweak sector; the ratio $\tan\beta=\frac{v_2}{v_1}$ of the VEVs of the two Higgs fields; the mixing angle $\alpha$ between the two scalar Higgs bosons $h$ and $H$; and the mixing angle $\theta$ between the pseudoscalar Higgs boson $A$ and the mediator $a$. 

The choice of the alignment limit ($\cos(\beta-\alpha)$=0) implies $m_h=125$~GeV and $v=246$~GeV.
The DM-mediator coupling is set to unity ($y_{\chi}=1.0$) without significant impact on the phenomenology of the model. The setting $\lambda_3=3$ is chosen to ensure the stability of the Higgs potential in the mass ranges of interested of the heavy Higgs bosons~\cite{ATLAS:2019wdu}. Furthermore, the choice $\lambda_{\textrm{P1}}$ =  $\lambda_{\textrm{P2}} = \lambda_3 = 3$ maximises the tri-linear couplings between the CP-even and CP-odd neutral states~\cite{ATLAS:2019wdu}. Finally, the choice $m_A = m_H = m_{H^{\pm}}$ ensures compatibility of the model predictions with flavour constraints~\cite{Bauer:2017ota} and additionally simplifies the phenomenology of the model~\cite{ATLAS:2019wdu}.

With these constraints, the remaining 2HMD+$a$ parameter space can be described by the following five parameters: $m_A$, $m_a$, $m_{\chi}$, $\sin\theta$, and $\tan\beta$. Representative benchmark scans of this parameter space have been defined by the LHC Dark Matter Working Group~\cite{LHCDarkMatterWorkingGroup:2018ufk} with the aim to highlight different aspects of the phenomenology of this benchmark model and the interplay between searches targeting different signal processes across this parameter space. Additional benchmark scans have been defined in Ref.~\cite{ATLAS:2HDMa_2021}.

The 2HDM+$a$ predicts a rich phenomenology with a diverse range of final states.
The dominant processes leading to final states with top quarks are shown in Figure~\ref{fig:SPS_Feyn},
along with the leading diagrams for the resonant production of an invisibly decaying mediator with a Higgs or $Z$ boson, leading to \met+$h$ and \met+$Z$ final states, respectively, which are among the most sensitive probes of the 2HDM+$a$. A full overview of the phenomenology of the 2HDM+$a$ can be found in Refs.~\cite{Bauer:2017ota,LHCDarkMatterWorkingGroup:2018ufk}.

\begin{figure}[h!]
\centering 
\includegraphics[width=0.24\textwidth]{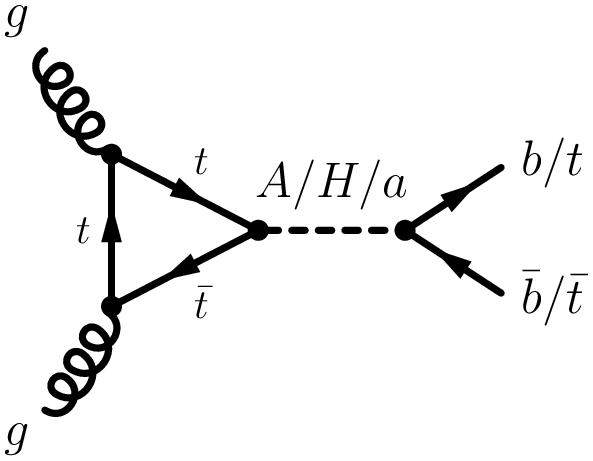}
\includegraphics[width=0.24\textwidth]{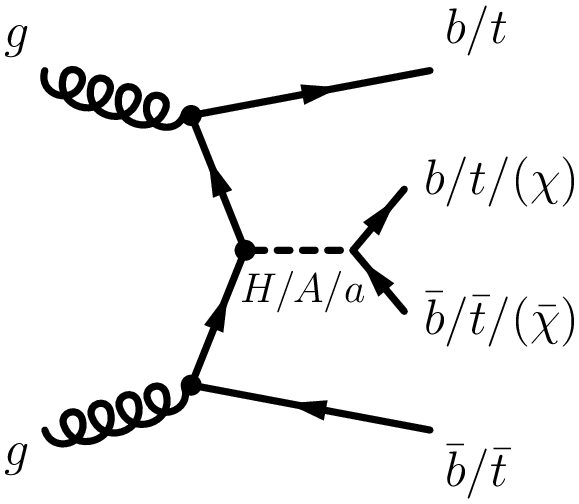}
\includegraphics[width=0.24\textwidth]{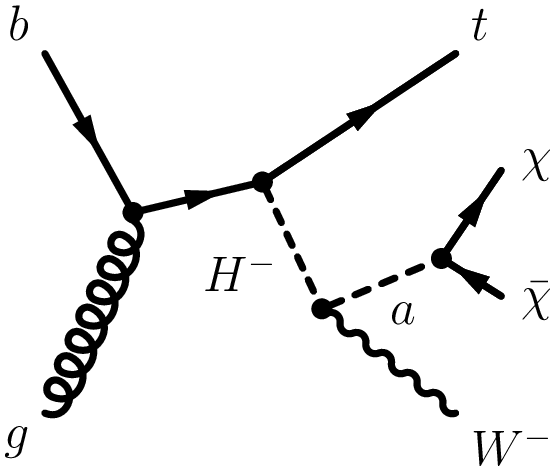}
\includegraphics[width=0.24\textwidth]{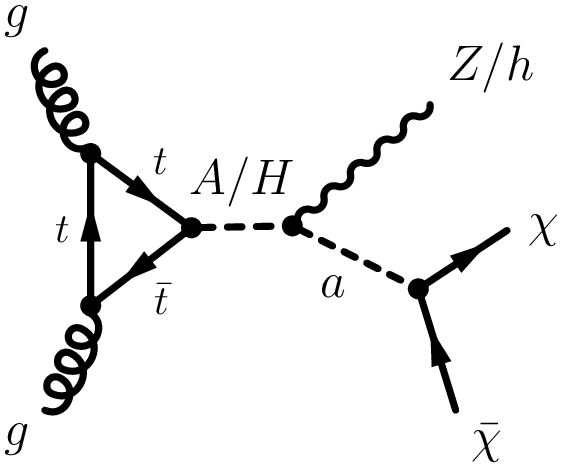}
\caption{Schematic representation of relevant production and decay modes with top quarks leading to either top quarks in the final state or \met+$h/Z$ signatures. From left to right: resonant production of a neutral scalar or pseudoscalar particle $H/A/a$ decaying to $t\bar{t}$ or $b\bar{b}$; associated production with $b\bar{b}$ or $t\bar{t}$ of a single $H/A/a$ decaying either visibly to heavy flavour or invisibly to DM; associated production of a top quark and a charged Higgs boson decaying to a $W$ boson and an invisibly decaying mediator $a$; resonant $A/H$ production with subsequent decay to a $Z/h$ boson and an invisibly decaying mediator $a$~\cite{ATLAS:2019wdu}.\label{fig:2HDM_Feyn}}
\end{figure}   

\subsection{EFT model of scalar dark energy}
\label{sec:DE_model}

Searches for DM signatures involving top quarks provide a powerful tool to probe models of scalar DE.
The first re-interpretation of DM searches in the context of DE, which relied on the analysis of 36~fb$^{-1}$ of LHC Run 2 data~\cite{ATLAS:2019wdu}, used an EFT implementation~\cite{Brax:2016did} of the Horndeski theories~\cite{Horndeski:1974} to describe DE production at the LHC~\cite{ATLAS:2019wdu}. 
The latter introduce a new scalar field, $\phi_{\mathrm{DE}}$, corresponding to DE, that couples to gravity.

The EFT model contains two classes of operators: 
operators that are invariant under a shift symmetry $\phi_{\mathrm{DE}} \rightarrow \phi_{\mathrm{DE} + \mathrm{constant}}$ and operators that break this symmetry. The former contain only derivative couplings of the DE field to SM fermions as direct Yukawa-type interactions break the shift symmetry.
The latter induce direct couplings of the DE field to the SM fermions, such as Yukawa-type interactions, and are subject to tight experimental constraints~\cite{Joyce:2014kja}.

Only shift-symmetric operators of the EFT model have been considered for the DE  re-interpretation of LHC DM searches~\cite{ATLAS:2019wdu}. The model under consideration contains nine such operators, 
$\mathcal{O}^{(d)}_i$, where $d$ denotes the dimensionality of the operator.
This leads to nine possible terms in the Lagrangian, 
each suppressed by powers of a characteristic energy scale $M_i^{d-4}$, according to the operator's dimensionality:
\begin{equation*}
\mathcal{L}=\mathcal{L}_{\mathrm{SM}}+\sum_{i=1}^9 c_i\mathcal{L}_i=\mathcal{L}_{\mathrm{SM}}+\sum_{i=1}^9 \frac{c_i}{M_i^{d-4}}\mathcal{O}^{(d)}_i,
\end{equation*} 
where the $c_i$ denote the Wilson coefficients. 

Only the phenomenology of the two leading, i.e. least suppressed, terms has been considered by the LHC experiments so far. These are of dimension eight and can be expressed in terms of the 
conformal anomaly, $T^{\nu}_{\nu}$ ($=m\bar{\psi}\psi$ for a Dirac field),
and the energy-momentum tensor of the SM Lagrangian $T^{\mu\nu}$
as follows:
\begin{eqnarray*}
\mathcal{L}_1&=&\frac{\partial_{\mu}\phi_{\mathrm{DE}}\partial^{\mu}\phi_{\mathrm{DE}}}{M_1^4}T^{\nu}_{\nu}\\
\mathcal{L}_2&=&\frac{\partial_{\mu}\phi_{\mathrm{DE}}\partial_{\nu}\phi_{\mathrm{DE}}}{M_2^4}T^{\mu\nu}.
\end{eqnarray*}

\begin{figure}[h!]
\centering 
\includegraphics[width=0.29\textwidth]{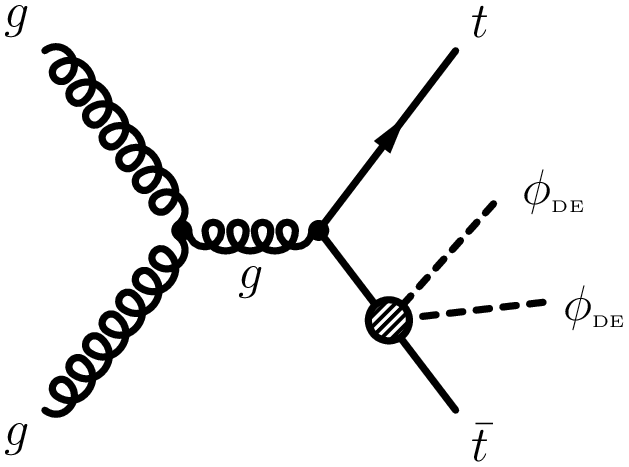}
\caption{Schematic representation of the leading process of DE production in association with a $t\bar{t}$ pair in an EFT model of scalar DE via the operator $\mathcal{L}_1$~\cite{ATLAS:2019wdu}.\label{fig:DE}}
\end{figure}

The coupling described by the first term, $\mathcal{L}_1$, 
is proportional to the mass of the SM fermions to which the DE field couples,
thus making collider signatures involving top quarks a sensitive probe of DE.
A schematic representation of DE production at the LHC via this operator is shown in Figure~\ref{fig:DE}.
It describes the radiation of a pair of DE particles off a final-state top quark from SM $t\bar{t}$ production, leading to a \met+$t\bar{t}$ signature.

The second operator, $\mathcal{L}_2$, involves derivatives of the SM fields and is therefore proportional to their momenta.
Final states involving high-momentum intermediate states, of which a DE pair is radiated off, provide the best sensitivity to this operator. At a hadron collider like the LHC, the most likely high-momentum intermediate state particles are hadronically interacting particles, such as gluons, leading to characteristic \met+jet signatures as the smoking-gun signatures for DE production.

Constraints on the EFT model of DE have been derived using searches for both \met+$t\bar{t}$ ($\mathcal{L}_1$ term) and \met+jet signatures~\cite{ATLAS:2019wdu} ($\mathcal{L}_2$ term). Only the former are discussed in this review. It should be noted that additional signatures, such as \met+$t$ production, are predicted based on the sub-leading operators. The exploration of these additional signatures and possible re-interpretations of further DM searches in the context of DE is left to future work.

\section{Experimental signatures}
\label{sec:signatures}

Searches for DM in $pp$ collisions involving single or multiple top quarks can be broadly split into two categories: Searches for large \met and searches for a DM mediator decaying into SM particles. Both classes rely on different analysis technique. Common to all searches is a detailed exploration of the top quark decay. Due to the almost diagonal structure of the CKM matrix and in particular $V_{tb}$ being close to one, the top quark decays almost 100\% of the time into a bottom quark and a $W$ boson. The $W$ boson itself decays with about 30\% probability into a charged lepton, i.e. an electron, muon, or tau, and the corresponding neutrino, or into two quarks otherwise. Similar to DM particles, neutrinos can only be inferred from missing transverse momentum in the detector. Events with two top quarks or with a single top quark and a $W$ boson are typically categorised in three orthogonal channels based on the lepton ($\ell = e,\mu$, including decays via $\tau$ leptons, i.e. $\tau \to $ e,$\tau \to \mu$) multiplicity in the final state.
0-lepton (0$\ell$) final states arise in events in which both $W$ bosons decay hadronically; 1-lepton (1$\ell$) final states arise in events in which one $W$ boson decays hadronically, the other leptonically; 2-lepton (2$\ell$) final states arise if both $W$ bosons decay leptonically.  

When top quarks recoil against significant \met or result from the decay of a very heavy resonance, top quarks are highly Lorentz boosted and their decay products become highly collimated. In the case of hadronic top-quark decays, this means that the particle showers from the three final-state quarks can no longer be reconstructed as three separate small-radius (small-$R$) jets (\textit{resolved decay)} but merge into a single large-radius (large-$R$) jet with characteristic substructure (\textit{merged decay}). Merged top-quark decays are identified using dedicated \textit{top tagging} algorithms.

\subsection{Final states with invisible decays}

\subsubsection{\met+$t$}
\label{sec:sig_inv_mett}

Searches for the production of large \met in association with a single top quark have been conducted by both the ATLAS~\cite{ATLAS-CONF-2022-036} and CMS \cite{CMS:2018gbj} Collaborations.

 

The ATLAS Collaboration has performed a \met+$t$ search targeting merged hadronic top-quark decays using 139~fb$^{-1}$ of $\sqrt{s}=13$~TeV $pp$ collision data~\cite{ATLAS-CONF-2022-036}.
Events are required to have \met>250~GeV and contain at least one large-$R$ (anti-$k_t$~\cite{Cacciari:2008gp} $R=1.0$) jet with transverse momentum $350 < p_T < 2500$~GeV and mass $40< m < 600$~GeV. Additionally, the selected jet must be identified as a top-quark candidate via a dedicated top-tagging algorithm~\cite{ATLAS:2018wis}, which relies on deep neural net (DNN) that uses jet kinematics and substructure variables as input~\cite{ATLAS:2018wis,ATLAS:DNNTopPerf}. The working point for the top tagging algorithm chosen for this analysis corresponds to a 50\% top tagging efficiency. 

Dedicated signal regions targeting resonant DM production via a colour-charged scalar mediator (Section~\ref{sec:SCC_model}) and non-resonant DM production via a vector mediator with a $V_{ut}$ coupling (Section~\ref{sec:VFC_model}) are defined based on the output score of XGBoost classifiers~\cite{XGBoost} that are trained on several event observables.
Control regions are defined to constrain the dominant backgrounds from $t\bar{t}$ and $V$+jets production.

A similar search has been performed by the CMS Collaboration~\cite{CMS:2018gbj}. Different from the ATLAS analysis the result is based on data recorded in 2016 only corresponding to an integrated luminosity of 36~\ifb. To identify the hadronically decaying top quark, CA15 jets are used. CA15 jets are clustered from particle flow candidates using the Cambridge–Aachen algorithm \cite{Cacciari:2008gp} with a distance parameter of 1.5.
The CA15 jets must have a transverse momentum $ p_T > 250~\gev $, $|\eta| < 2.4$ and an invariant mass of $110~\gev < m < 210~\gev$. Furthermore, several substructure observables, like the N-subjettiness \cite{Thaler:2010tr} or  so-called energy-correlation functions \cite{Larkoski:2013eya,Moult:2016cvt} are combined in a boosted decision tree (BDT)~\cite{Hocker:2007ht} to distinguish top quark jets from the hadronisation products of single light quarks or gluons. At 50\% signal efficiency, the BDT background acceptance is 4.7\%. The dominant backgrounds from \ttbar and single vector bosons (Z, W, $\gamma$) are constraint from dedicated control regions. The signal is probed in distributions of missing transverse energy \met considering two signal regions which correspond to a BDT output between 0.1 and 0.45 and above 0.45 respectively.  

The summary plots for the benchmark model with a colour-charged scalar mediator in Section~\ref{sec:SCC_results}, which show the interplay between the \met+$t$ and same-sign $tt$ (Section~\ref{sec:sig_vis_ss_tt}) searches, are based on an earlier search of the ATLAS Collaboration using 36~fb$^{-1}$ of $\sqrt{s}=13$~TeV $pp$ collision~\cite{ATLAS:2018cjd}. This analysis statistically combines the results from two orthogonal channels, targeting semi-leptonic and hadronic top-quark decays, respectively.

\subsubsection{\met+$tW$ and \met+$tj$}
\label{sec:sig_inv_mettWtj}

Like the \met+$t$ searches described in Section~\ref{sec:sig_inv_mett}, searches for \met+$tW$ target events with single top quarks produced in association with large \met but additionally require the existence of a second visible object. This can be either a $W$ boson or a hadronic jet. The resulting signatures are referred to as \met+$tW$ and \met+$tj$, respectively. It should be noted that searches in these final states are not orthogonal to the \met+$t$ searches discussed in Section~\ref{sec:sig_inv_mett} as the latter do not veto the presence of additional visible objects in the event and hence implicitly include \met+$tj$ and \met+$tW$ signatures. 

While \met+$t$ searches are traditionally used to constrain resonant DM production via a colour-charged scalar mediator and non-resonant DM production via a vector mediator with a flavour-violating $V_{ut}$ coupling, as explained in Section~\ref{sec:sig_inv_mett}, \met+$tW$ searches in particular are used to probe the 2HDM+$a$ (Section~\ref{sec:2HDMa_model})
and more recently also simplified models with a scalar or pseudoscalar mediator (Section~\ref{sec:SPS_model}).

Simplified models with a scalar or pseudoscalar mediator predict both \met+$tW$ and \met+$tj$ production,  as illustrated by the two right-most diagrams in Figure~\ref{fig:SPS_Feyn}. The corresponding signal cross-sections are, up to mediator masses of 200~\gev, smaller than those of the dominant \met+$t\bar{t}$ production mode discussed in Section~\ref{sec:sig_inv_ttmet}.
Therefore, \met+$tW$ and \met+$tj$ searches have not been used to constrain these simplified models by the ATLAS Collaboration. However, with the increased sensitivity of recent searches, single top associated production becomes more and more relevant and a first search including \met+$tW$  and  \met+$t\bar{t}$ signatures has been performed by the CMS Collaboration~\cite{CMS:2019zzl} as further discussed in Section~\ref{sec:sig_inv_mettttwtj}.

\met+$tW$ and \met+$tj$ production is also predicted in the 2HDM+$a$. Compared to simplified models with a single (pseudo)scalar mediator, this model contains additional production modes, illustrated for example by the third diagram in Figure~\ref{fig:2HDM_Feyn}, which lead to higher predicted signal cross-sections for \met+$tW$ and \met+$tj$ production.
A search for \met+$tW$ and \met+$tj$ signatures, optimised specifically for 2HDM+$a$ signal processes, has been conducted by the ATLAS Collaboration~\cite{ATLAS:2020yzc} using 139~fb$^{-1}$ of $\sqrt{s}=13$~TeV $pp$ collision data.
The search considers events with one or two leptons ($e$,$\mu$), at least one $b$-tagged jet, and significant \met in three orthogonal categories. Two of them target \met+$tW$ production in final states with one or two leptons, while  
the third channel targets \met+$tj$ production in final states with exactly one lepton.
The search has been extended in the context of a preliminary analysis of the same dataset~\cite{ATLAS:METplusTopW} to include events with highly energetic $W$ boson decays in final states with zero leptons or one lepton. These provide additional sensitivity for large masses of the charged Higgs bosons. The newly added zero- and improved one-lepton channels are statistically combined with the two-lepton channel of Ref.~\cite{ATLAS:2020yzc}.

\subsubsection{\met+$t\bar{t}$}
\label{sec:sig_inv_ttmet}

Searches for DM or DE production in association with a $t\bar{t}$ pair target final states characterised by sizeable \met and the presence of the $t\bar{t}$ decay products.

The CMS Collaboration has released a search for DM in association with a $t\bar{t}$ pair using 137~\ifb of data recorded at \sqrts = 13~\tev between 2016 and 2018~\cite{CMS:2021eha}. The analysis combines previous searches in final states with 0~\cite{cms_stop_0l}, 1~\cite{cms_stop_1l} or 2~\cite{cms_stop_2l} leptons. While the primary target of the analyses is stop quark production, a re-interpretation of the combined result in a simplified DM model with scalar mediators is provided.


Central feature of the analysis in the $0$-lepton channel is an advanced jet-tagging algorithm identifying hadronically decaying top quarks and $W$ bosons with low and high Lorentz-boost. For the highly Lorentz-boosted regime, the DeepAK8 algorithm \cite{cms_deepak8} is used whereas in the resolved regime the DeepResolved algorithm \cite{cms_stop_1l} is explored to tag top quarks in the intermediate transverse momentum range from 150 to 450 \gev.
The analysis includes a total of 183 non-overlapping signal regions. The contribution of each SM background process is estimated through measurements of event rates in dedicated background control samples that are translated to predicted event counts in the corresponding signal region with the aid of MC simulation.


The key requirements in the $1$-lepton channel are exactly one lepton and $\met > 250~\gev$. Moreover, the transverse mass computed from the lepton and the missing momentum is required to be larger than 150 \gev to reduce the dominant  background from SM \ttbar and $W$+jets production, for which the transverse mass has a natural cutoff at the mass of the $W$ boson. The SM production of dileptonic \ttbar  events, where one of the leptons is lost, is the largest remaining background. It is  estimated through a set of dedicated control regions and reduced by using the modified topness variable~\cite{cms_stop_1l}. The $1$-lepton channel also exploits the jet tagging algorithms used in the $0$-lepton channel, to identify hadronic top quark decays. 
In order to enhance the sensitivity to different signal scenarios, including the case of small missing transverse momentum, events are categorised into a total of 39 non-overlapping signal regions. 


The search in the $2$-lepton channel explores orthogonal signal regions based on the flavour of the leptons and three characteristic observables: The so-called missing transverse momentum significance~\cite{cms_metsig} and two specific definitions of the stransverse mass~\cite{cms_stop_2l,cms_stransversemass}. The \met significance is given by the ratio of the \met over its resolution and it is particularly  powerful to suppress events where detector effects and misreconstruction of particles from pileup interactions are the main source of missing transverse momentum. The key feature of the stransverse mass using leptons (lepton and b-quark jets) is that it retains a kinematic endpoint at the $W$-boson (top-quark) mass for SM background events from the leptonic decays of two $W$ bosons (top quarks). The dominant backgrounds arise from $t\bar{t}$ and $t\bar{t}+Z$ production as well as single-top quark production in the $Wt$ channel. After a veto of the $Z$-boson mass window, i.e. $|m_{\ell\ell} - m_Z|> 15$~GeV, Drell-Yan production represents only a minor source of background. 

A similar search using 139~\ifb of LHC data has been released by the ATLAS Collaboration exploring separately the 0-lepton~\cite{ATLAS:2020dsf}, 1-lepton~\cite{ATLAS:2020xzu}, and 2-lepton~\cite{ATLAS:2021hza} channels. 
All three final states have been combined afterwards into a single result~\cite{ATLAS:METplustt}. In this context, the (0$\ell$) channel search has been further optimised through an improved selection of triggers targeting $b$-jets. Searches for \met+$tW$ (Section~\ref{sec:sig_inv_mettWtj}) production have not been included in this combination as their datasets are not orthogonal to those in the \met+$t\bar{t}$ by construction. Including them in a statistical combination is left to future publications.
While by now the \met+$t\bar{t}$ searches discussed above have been interpreted in simplified models with a scalar or pseudoscalar mediator only, see Section~\ref{sec:SPS_results}, earlier searches, based on smaller datasets, have already been used to constrain a 2HDM with a pseudoscalar mediator (Section~\ref{sec:2HDMa_results}) and a model of scalar DE (Section~\ref{sec:DE_model}). 

\subsubsection{\met+$tW$, \met+$tj$ and \met+\ttbar}
\label{sec:sig_inv_mettttwtj}
A first result exploring topologies of single top quark and top-quark pair associated production has been released by the CMS Collaboration~\cite{cms_exo_18_010}. The analysis is using 36~\ifb of data recorded in 2016 at 13~\tev and combines multiple selection categories in final states with 0 or 1 lepton. In the 1 lepton channel, dominant background is suppressed using a similar strategy as the one discussed in Section~\ref{sec:sig_inv_ttmet}, while in the 0 lepton channel, dominant background is reduced by a cut on the missing transverse energy, the ratio of the leading jet transverse momentum over the total hadronic transverse energy in the event, and the minimum opening angle between the missing transverse energy and the two leading jets. To enhance the sensitivity to single top quark associated production, events are separated according to the number of identified b-quark jets. Events with a single b-tagged jet are further split into events with a central or forward jet. The categorization in terms of forward jets allows a further enhancement of t/$\bar{\rm t}$+DM t-channel events. This production mode leads to final states with one top quark and an additional jet, which tends to be in the forward region of the detector, while the additionally produced b quark is typically low in transverse momentum and therefore not reconstructed.
Key observable of this search is the \met spectrum explored in a combined fit to different orthogonal signal regions. Overall, data are found to be in good agreement with the expected SM background. Due to the combination of single top quark and \ttbar associated production, this analysis was able to derive the most stringent limits from LHC data on spin-0 mediators at that time.  

\subsection{Final states without invisible decays}

\subsubsection{Same-sign $tt$}
\label{sec:sig_vis_ss_tt}

Events with a same-sign $tt$ pair are identified via the leptonic decays of the $W$ bosons from the two top quarks.
They are required to contain two same-sign charged leptons, at least one $b$-jet, and significant \met from the two neutrinos resulting from the leptonic $W$ boson decays.

A search in same-sign $tt$ events has been conducted by the ATLAS Collaboration, using 36~\ifb of $\sqrt{s}=13$~TeV data~\cite{ATLAS:2018alq}.
The signal region of this search is defined by requiring the presence of two positively charged leptons ($e$,$\mu$) and at least one $b$-jet. Additionally, the scalar sum of the transverse momenta of all selected objects in the event, $H_T$ is required to be significant ($H_T > 750$~GeV). Further requirements on the \met and the angular separation of the two leptons are imposed. The signal region is split into three orthogonal channels based on the lepton flavour ($ee$, $e\mu$, $\mu\mu$).
The main backgrounds of this search are estimated using MC simulation, while the sub-dominant background from fake leptons is estimated using data-driven techniques.

\subsubsection{$t\bar{t}$}
\label{sec:sig_vis_ttbar}


A search for resonant \ttbar production in the $0\ell$ channel has been conducted by the ATLAS Collaboration using 139~\ifb of $\sqrt{s}=13$~TeV data~\cite{ATLAS:2020lks}. This search targets heavy vector and axial-vector resonances (including DM mediators) with masses $>1.4$~TeV, resulting in two merged top-quark decays. Merged top-quark decays are identified using a deep-neural net (DNN) based top tagger trained on the distributions of various characteristic jet and jet substructure variables to distinguish top-quark from light-quark and gluon initiated jets.
SM \ttbar production constitutes the main, irreducible background to this search, followed by strong multi-jet production. The background spectrum is derived from data by fitting a smoothly falling function to the reconstructed $m_{\ttbar}$ distribution, similar to the approach classically chosen in di-jet resonance searches.

A larger range of resonance masses has been probed by a search for resonant \ttbar production in the $1\ell$ channel, conducted by the ATLAS Collaboration on 36~\ifb of $\sqrt{s}=13$~\tev data~\cite{ATLAS:2018rvc}. This search targets both \textit{merged} and \textit{resolved} hadronic top-quark decays and is sensitive to resonance masses just above the \ttbar kinematic threshold ($>2\mtop$). The main, irreducible background from SM \ttbar production, as well as most other, smaller backgrounds, are estimated using MC simulation. Data-driven corrections are applied to the MC simulation of the $W$+jets background. The small background from strong multi-jet production is estimated with a fully data-driven approach.

A first search for heavy spin-1 resonances combining final states with 0, 1 and 2 leptons has been performed by the CMS Collaboration using data recorded at \sqrts=13~\tev and corresponding to a total integrated luminosity of 35.9~\ifb~\cite{CMS:2018rkg}. 
The analysis utilises reconstruction techniques that are optimised for top quarks with high Lorentz boosts, which requires the use of non-isolated leptons partially overlapping with $b$-quark jets and jet substructure techniques for top-quark tagging. Except for the QCD multijet background in the 0-lepton channel, the shapes of all backgrounds are estimated from MC simulation. The signal strength is extracted from the distributions of the reconstructed invariant mass of the top quark pair for the 0- and 1-lepton channels and from the sum of missing transverse energy and the transverse momenta of all jets and leptons in the 2-lepton channel. 


Interference effects between the resonant signal and background processes are not taken into account in the searches discussed above as they are irrelevant for spin-1 and spin-2 particles. However, this is not true for scalar and pseudoscalar resonances, such as additional heavy Higgs bosons, which are produced from $gg$ initial states via heavy quark loops.
The process $gg\rightarrow A/H \rightarrow \ttbar$ interferes strongly with the irreducible background from SM \ttbar production, which is dominated by $gg\rightarrow \ttbar$.
Interference effects significantly distort the resonance lineshape from a Breit-Wigner peak to a characteristic peak-dip or even more complicated structures.
The treatment of these effects is non-trivial and requires dedicated analysis methods, in particular in the statistical analysis.
Searches for heavy scalars and pseudoscalars have been conducted by both the ATLAS~\cite{ATLAS:2017snw} and CMS Collaborations~\cite{CMS:2019pzc}
in the $1\ell$ and $1\ell+2\ell$ channels, respectively. 
These searches are sensitive to the production of scalar and pseudoscalar DM mediators. However, due to the strong model-dependence of the interference patterns, no dedicated interpretation of these results in the context of DM models exists to date. An approximate re-interpretation of the results in Ref.~\cite{ATLAS:2017snw} in the context of the 2HDM+$a$ (Section~\ref{sec:2HDMa_model}) can be found in Ref.~\cite{Bauer:2017ota}.

\subsubsection{$t\bar{t}t\bar{t}$}
\label{sec:sig_vis_4top}

Final states with four top quarks ($\ttbar\ttbar$) can arise from non-resonant processes predicted in the SM but are also predicted in BSM models allowing for the associated production of a heavy BSM resonance, which subsequently decays to \ttbar, with a \ttbar pair.
Four-top final states are particularly relevant in searches for heavy scalars and pseudoscalars, as the signal-background interference is negligible for associated production with \ttbar compared to loop-induced production from $gg$ initial states (Section~\ref{sec:sig_vis_ttbar}). It should be noted, though that the production cross-section for associated production is significantly lower than for loop-induced production.

Four-top final states are characterised by a high object multiplicity. Orthogonal signal regions can be defined based on the multiplicity of leptons ($e,\mu$) in the final state, which corresponds to the number of top quarks with a leptonically decaying $W$ boson.

The ATLAS Collaboration has recently found evidence ($4.3~\sigma$ observed, $2.4~\sigma$ expected significance) for four-top quark production in a search focusing on the multi-lepton final state conducted on 139~\ifb of $\sqrt{s}=13$~TeV $pp$ collision data~\cite{ATLAS:2020hpj}. The result is consistent with the SM prediction for four-top production within $1.7\sigma$. 
A subsequent dedicated search for BSM four-top production on the same dataset specifically targets \ttbar associated production of heavy scalar or pseudoscalar Higgs bosons $A/H$ decaying to \ttbar (\ttbar $A/H\rightarrow \ttbar\ttbar$)~\cite{ATLAS:2022rws}.
It is based on and extends the analysis strategy of Ref.~\cite{ATLAS:2020hpj} to increase the sensitivity to $A/H$ production.
In both the SM and BSM searches, events are required to contain either a same-sign lepton pair or at least three leptons.
A multivariate discriminant based on a Boosted Decision Tree (BDT) is used to separate between SM four-top production and other background processes, using event-level information such as jet and $b$-jet multiplicity as well as additional kinematic variables.
The BSM search relies on a second BDT to subsequently distinguish between BSM and SM four-top production. This second BDT is parameterised as a function of the mass of the heavy Higgs boson by introducing the mass as a labelled input in the training~\cite{Baldi:2016fzo}.
The main, irreducible backgrounds arise from associated production of a \ttbar pair with a boson and additional jets (\ttbar+$W$+jets, \ttbar+$Z$+jets, \ttbar+$h$+jets). They are estimated using MC simulations with additional data-driven corrections applied in the case of \ttbar+$W$+jets production. Smaller, reducible backgrounds arise mostly from \ttbar+jets and $tW$+jets production with mis-identified charge or fake/non-prompt leptons. These smaller backgrounds are estimated from data using dedicated control regions.
No significant excess of events over the SM prediction is observed in the BSM four-top search and the results are interpreted in the context of a type-II 2HDM. No dedicated interpretation in the context of DM models has been performed. The constraints on the type-II 2HDM with $m_A=m_H$, however, indicate that this search can improve upon the current four-top constraints on the 2HDM+$a$ parameter space included in the latest 2HDM+$a$ summary plots of Ref.~\cite{ATLAS:DMSum} (Section~\ref{sec:2HDMa_results}), which are based on a search in the single-lepton channel using 36~\ifb of $\sqrt{s}=13$~TeV data~\cite{ATLAS:2017oes}. 

The CMS Collaboration has reported an observed (expected) significance for $\ttbar\ttbar$ of $2.6~\sigma$ ($2.7~\sigma$) in the multi-lepton channel using 137~\ifb of $\sqrt{s}=13$~TeV $pp$ collision data~\cite{CMS:2019rvj}. The search relies on a new multivariate classifier to
maximize the sensitivity to the SM $\ttbar\ttbar$ signal. As in the equivalent ATLAS search, the main backgrounds from $\ttbar$+boson+jets production are estimated using MC simulations. Data-driven corrections are applied in the cases of \ttbar+$W$+jets and \ttbar+$Z$+jets production.
Backgrounds arising from charge mis-identification or fake/non-prompt leptons are estimated from data. This result has been used to constrain scalar and pseudoscalar production in 2HDMs as well as in the simplified DM model with a scalar or pseudoscalar mediator (Section~\ref{sec:SPS_model}). No dedicated interpretation for the 2HDM+$a$ is available, although the constraints on type-II 2HDMs suggest that the search will also constrain the 2HDM+$a$ parameter space.

The searches described above have been optimised for non-resonant $\ttbar\ttbar$ production and/or production of heavy scalar or pseudoscalar resonances, including resonance masses below 1 TeV.
An additional search targeting top-philic vector and axial-vector ($Z'$) resonances with masses $>1$~TeV has been conducted by the ATLAS Collaboration. The preliminary result relies on 139~$\ifb$ of $\sqrt{s}=13$~TeV data~\cite{ATLAS:ttZp}. Unlike other searches in the $\ttbar\ttbar$ final state, this search is designed to reconstruct the BSM resonance explicitly from a pair of re-clustered jets identified as merged top quarks.
The results can in principle be used to constrain purely top-philic vector or axial-vector mediators to which classic \ttbar resonance searches, which assume $Z'$ production from light-quark or gluon initial states (Section~\ref{sec:sig_vis_ttbar}) may not be sensitive. A dedicated interpretation of this search in the context of DM models is left to future work.

\subsubsection{$tbH^{\pm}(tb)$}
\label{sec:sig_vis_tbtb}

Final states with two top and two bottom quarks are sensitive to the associated production of a charged Higgs boson $H^{\pm}$ with a top and a bottom quark ($tb$) and its subsequent decay to $tb$. 

The ATLAS Collaboration has published a search for $tbH^{\pm}(tb)$ production using 139~\ifb of $\sqrt{s}=13$~TeV data~\cite{ATLAS:2021upq}. It targets charged Higgs boson masses in the range 0.2–2.0 TeV.
Events are required to contain exactly one electron or muon 
to suppress the large backgrounds from strong multi(-$b$)-jet production.
The selected events are further classified according to the number of reconstructed jets and the number of $b$-jets among them.
A neural network is used to enhance the separation between signal and background.
The dominant background for this search is composed of \ttbar jets events as well as single-top production in the $Wt$ channel. The backgrounds are modelled using MC simulations with additional data-driven corrections derived in a dedicated control region.

A search for charged Higgs bosons decaying into a top and a bottom quark in
the 0-lepton final state has been performed by the CMS Collaboration using proton-proton collision at \sqrts = 13 \tev from 2016 ~\cite{CMS:2020imj}. Two different scenarios have been studied, the associated production with a top and bottom quark and the $s$-channel production of a charged Higgs. The results are combined with a search in final states with one or two leptons~\cite{CMS:2019rlz}. For production in association with a top quark, upper limits at the 95\% confidence level on the charged Higgs production cross section and branching fraction of 9.25 to 0.005 pb are obtained for charged Higgs masses in the range of 0.2 to 3 \tev. 
While there is no DM interpretation of the result by the CMS Collaboration, the result from ATLAS was interpreted in a 2HDM+$a$ scenario, as further detailed in Section~\ref{sec:2HDMa_results}.

\section{Results}
\label{sec:results}

\subsection{Vector and axial-vector mediators}

\subsubsection{Flavour-conserving interaction}
\label{sec:AVV_results}

Strong constraints on visible decays of the axial-vector (Figure~\ref{fig:AVV_limits}) or vector (Figure~\ref{fig:VV_limits}) mediator $m_{Z'}$ are obtained from a variety of resonance and related searches that probe mediator masses in the range between 50~GeV~\cite{CMS:DMSum} and 5000~GeV~\cite{ATLAS:DMSum}. 

The latest constraints on axial-vector mediators released by the ATLAS Collaboration and based on data from $pp$ collisions at $\sqrt{s}=13$~TeV are shown in Figure~\ref{fig:AVV_limits}. The coupling of the mediator to leptons is set to zero ($g_{\ell}=0$), while the coupling to DM is set to unity ($g_{\chi}=1.0$) and the DM mass is taken to be 10~TeV to kinematically suppress invisible mediator decays and highlight the interplay of constraints on visible mediator decays.

In the high mediator mass range, the main sensitivity comes from two searches for di-jet resonances, referred to as \textit{di-jet} and \textit{di-jet angular}. The former aims to identify local resonant enhancements in the di-jet invariant mass spectrum and targets narrow mediator widths. The latter, for which no results on the full LHC Run 2 dataset are available, relies on the di-jet angular separation to identify broader mediator widths that cannot be probed by the search in the invariant mass spectrum.
Neither of the searches imposes quark-flavour specific selection requirements and hence are sensitive to all possible hadronic decays of the mediator.

Searches for $t\bar{t}$ resonances, which rely on top-quark identification algorithms to identify specifically the decays of the mediator to top quarks, have a slightly lower expected sensitivity to the coupling $g_q$ than di-jet searches, although the observed limit is stronger than that from the di-jet search in some small regions of the mediator mass where the di-jet observed limit fluctuates upward.
The use of top-quark identification allows for a stronger suppression of SM backgrounds compared to di-jet and also di-$b$-jet searches, in particular the background from strong multi-jet production. This effect partially compensates the disadvantage of probing only roughly $\frac{1}{6}$ of the hadronic mediator decays.

In Figure~\ref{fig:VV_limits}, constraints on vector mediators in the plane of the DM and the mediator mass from the CMS Collaboration~\cite{CMS:DMSum} are shown. Different from Figure~\ref{fig:AVV_limits}, results from visible and invisible decays are summarised. While searches with invisible final states are only possible when the mediator mass is about twice the DM mass, the sensitivity of searches for visible decays only depends on the DM mass through the width of the mediator. When the decay channel to DM particles opens up, the width of the mediator increases and resonant searches become less sensitive.
The best sensitivity to vector mediators from \ptmiss searches is provided by DM searches with initial state radiation either from a gluon/quark jet or from the hadronic decay of a vector boson~\cite{CMS:2021far}. Searches with visible final states achieve best sensitivity down to 50~\gev when looking for a large radius jet that recoils against the mediator~\cite{CMS:2019emo}. At high mass, the strongest constraints are obtained from di-jet searches~\cite{CMS:2019gwf}. The searches discussed in Section~\ref{sec:sig_vis_ttbar} probing vector mediators decaying into \ttbar are not shown as no dedicated interpretation of these results where performed in models of DM by the CMS Collaboration. However, the interpretation of the searches in generic vector particle models show comparable sensitivity between the results released by the ATLAS and CMS Collaborations. 

\begin{figure}[p]
\centering
\includegraphics[width=0.7\textwidth]{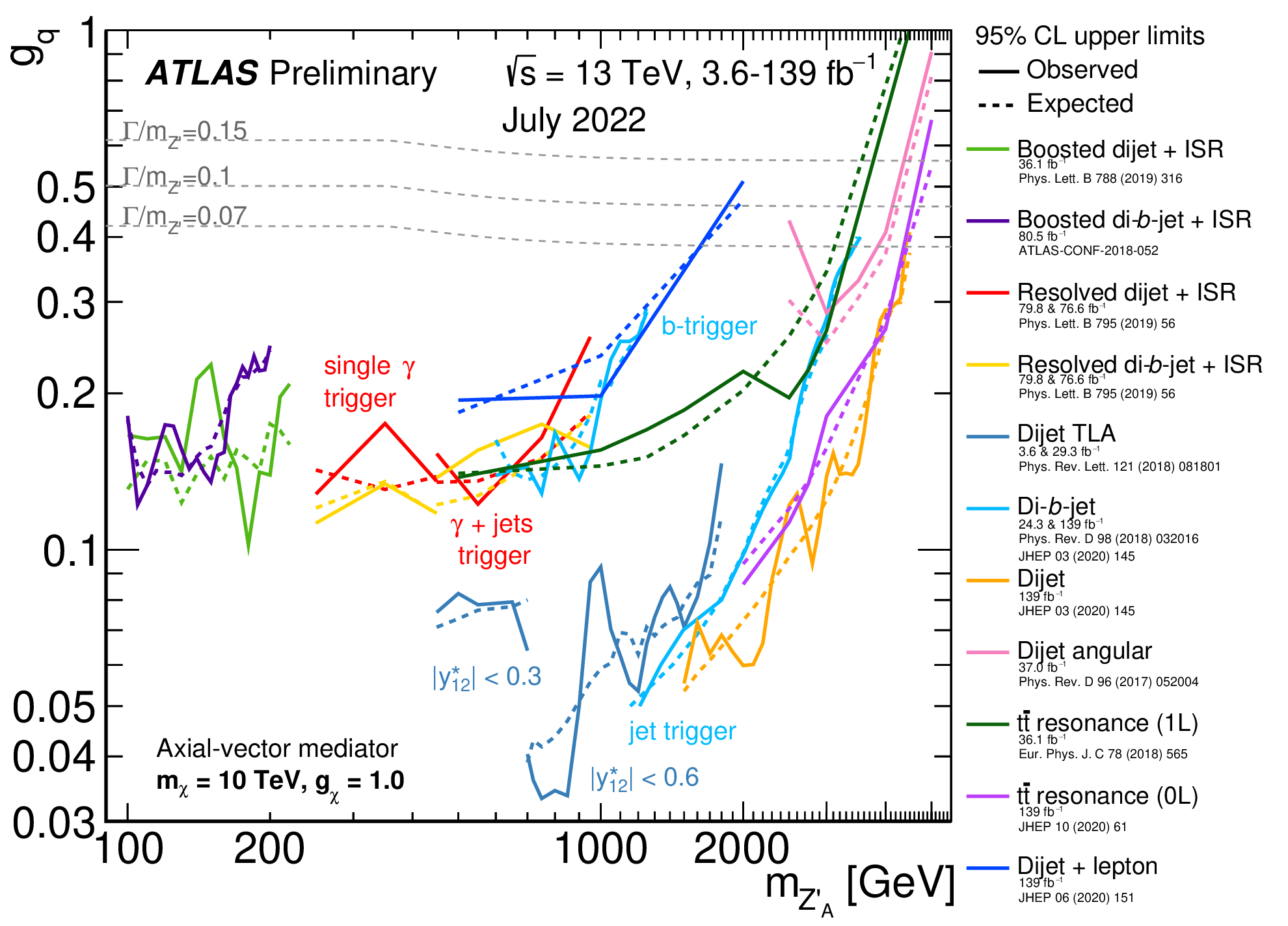}
\caption{Upper limits at 95\% CL on the coupling $g_q$ of the mediator to quarks in a simplified model with a vector or axial-vector mediator obtained from different types of resonance searches using data from $pp$ collisions at $\sqrt{s}=13$~TeV. The DM mass is $m_{\chi}=10$~TeV and its coupling to the mediator $g_{\chi}=1$~\cite{ATLAS:DMSum}.\label{fig:AVV_limits}}
\end{figure}  

\begin{figure}[p]
\centering
\includegraphics[width=0.7\textwidth]{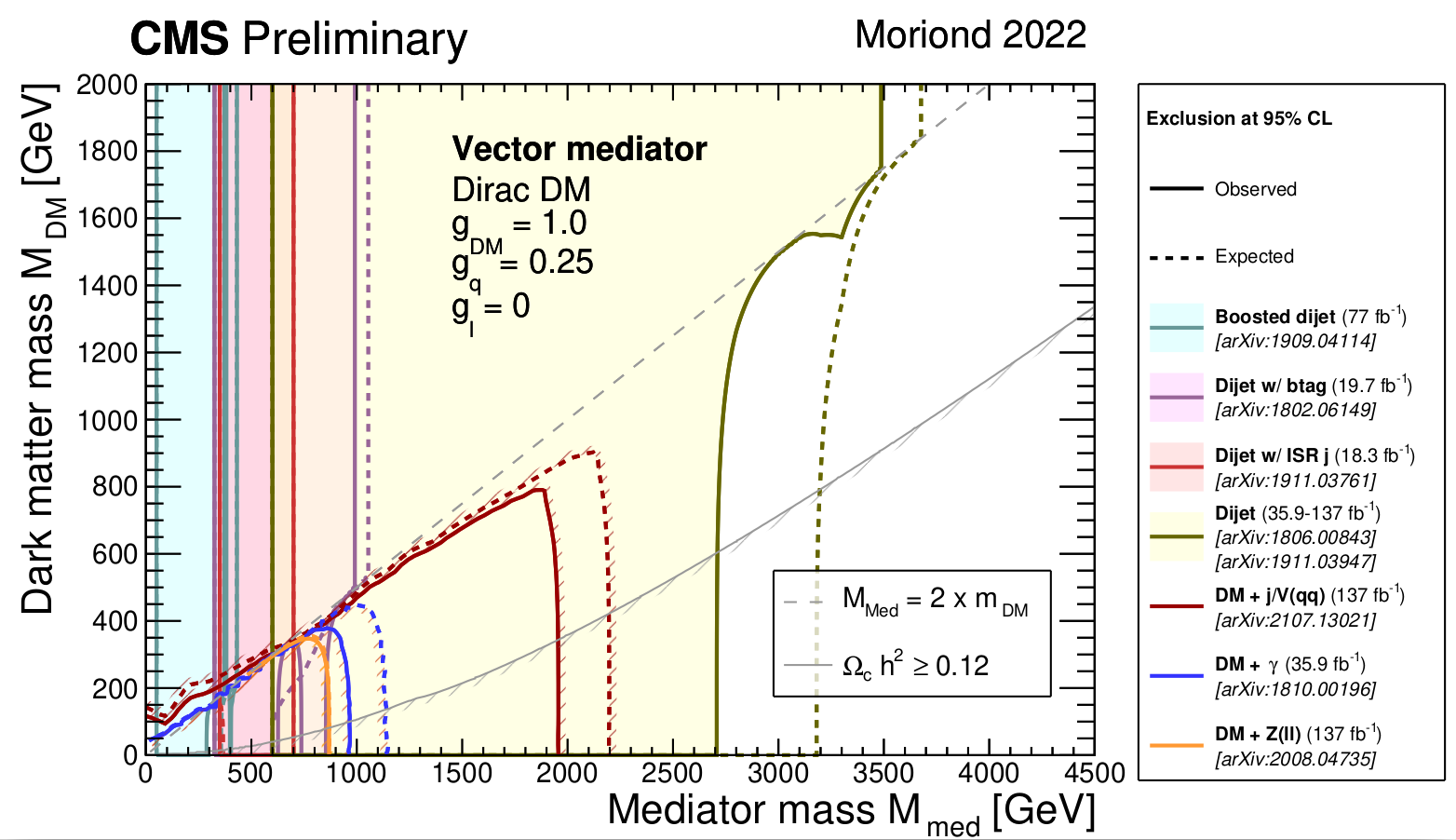}
\caption{95\% CL observed and expected exclusion regions on vector mediators in the DM-mediator mass plane from searches with visible and invisible final states released by the CMS Collaboration~\cite{CMS:DMSum}. Exclusions are computed for a lepto-phobic scenario with $g_l=0$, a universal quark coupling of $g_q = 0.25$ and a DM coupling of $g_{\rm DM} = 1.0$.\label{fig:VV_limits}}
\end{figure}  

\subsubsection{Flavour-changing interaction}
\label{sec:VFC_results}

The strongest constraints on the VFC model are obtained from searches targeting same-sign $tt$ and \met+$t$ production on 36~fb$^{-1}$ of $pp$ collision data~\cite{ATLAS:2019wdu}.
Results for two representative parameter planes are shown in Figure~\ref{fig:VFC_limits}.

The left plot of Figure~\ref{fig:VFC_limits} shows a scan in the mediator mass versus the flavour-changing coupling $g_{ut}$ while fixing the remaining two parameters at $m_{\chi}=1$~GeV and $g_{\chi}=1$.
The \met+$t$ search provides stronger constraints on $g_{ut}$ at lower mediator masses, excluding $g_{ut}$ down to 0.07 at 1~TeV, while the same-sign $tt$ search is more sensitive for mediator masses > 1.6 TeV, still excluding $g_{ut}>0.3$ at 3~TeV.
Mediator masses below 1~\tev have been probed by the CMS Collaboration at \sqrts = 13~\tev and are shown in Figure~\ref{fig:VFC_limits_CMS}. The \met+$t$ search discussed in Section~\ref{sec:sig_inv_mett} is able to exclude couplings as low as 0.03 for mediator masses of 200~\gev.
The right plot of Figure~\ref{fig:VFC_limits} shows a scan in the invisible branching ratio of the mediator $\mathcal{BR}(\chi \chi)$ and the coupling $g_{ut}$. The constraints derived from the same-sign $tt$ search exhibits only a weak dependence on $\mathcal{BR}(\chi \chi)$ due to the fact that the sensitivity of this process is dominated by the $t$-channel exchange of the mediator (middle and right diagrams in Figure~\ref{fig:VFC_Feyn}). This process is only indirectly sensitive to $g_{\chi}$ through the total width of the mediator in the $t$-channel exchange. The same-sign $tt$ analysis hence dominates the sensitivity at low values of $g_{\chi}$ (and hence low values of $\mathcal{BR}(\chi \chi)$), while the \met+$t$ analysis dominates the sensitivity at large values of $\mathcal{BR}(\chi \chi)$, excluding $g_{ut}$ down to almost 0.06 at $\mathcal{BR}(\chi \chi) = 1$.
\begin{figure}[h!]
\centering
\includegraphics[width=0.49\textwidth]{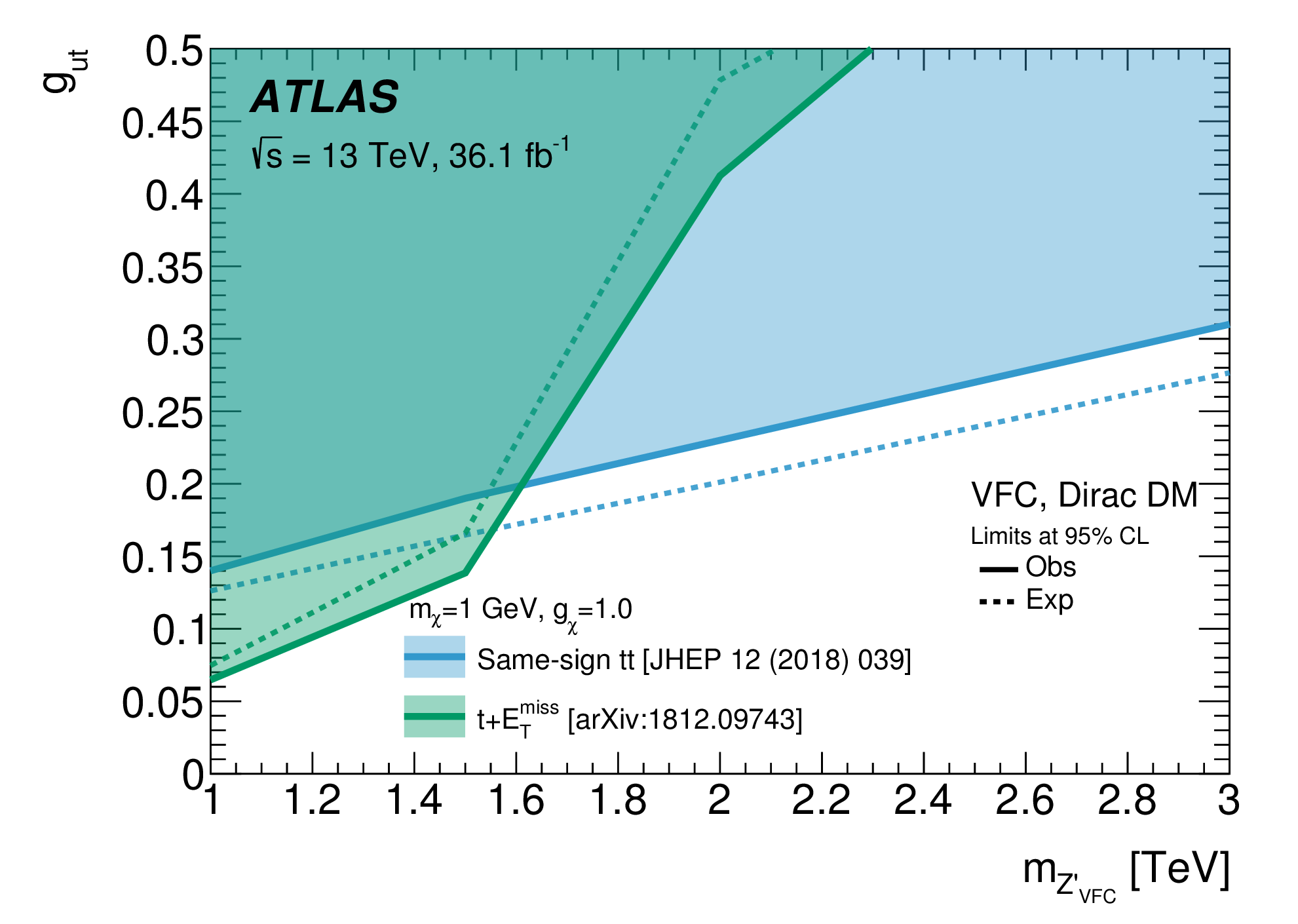}
\includegraphics[width=0.49\textwidth]{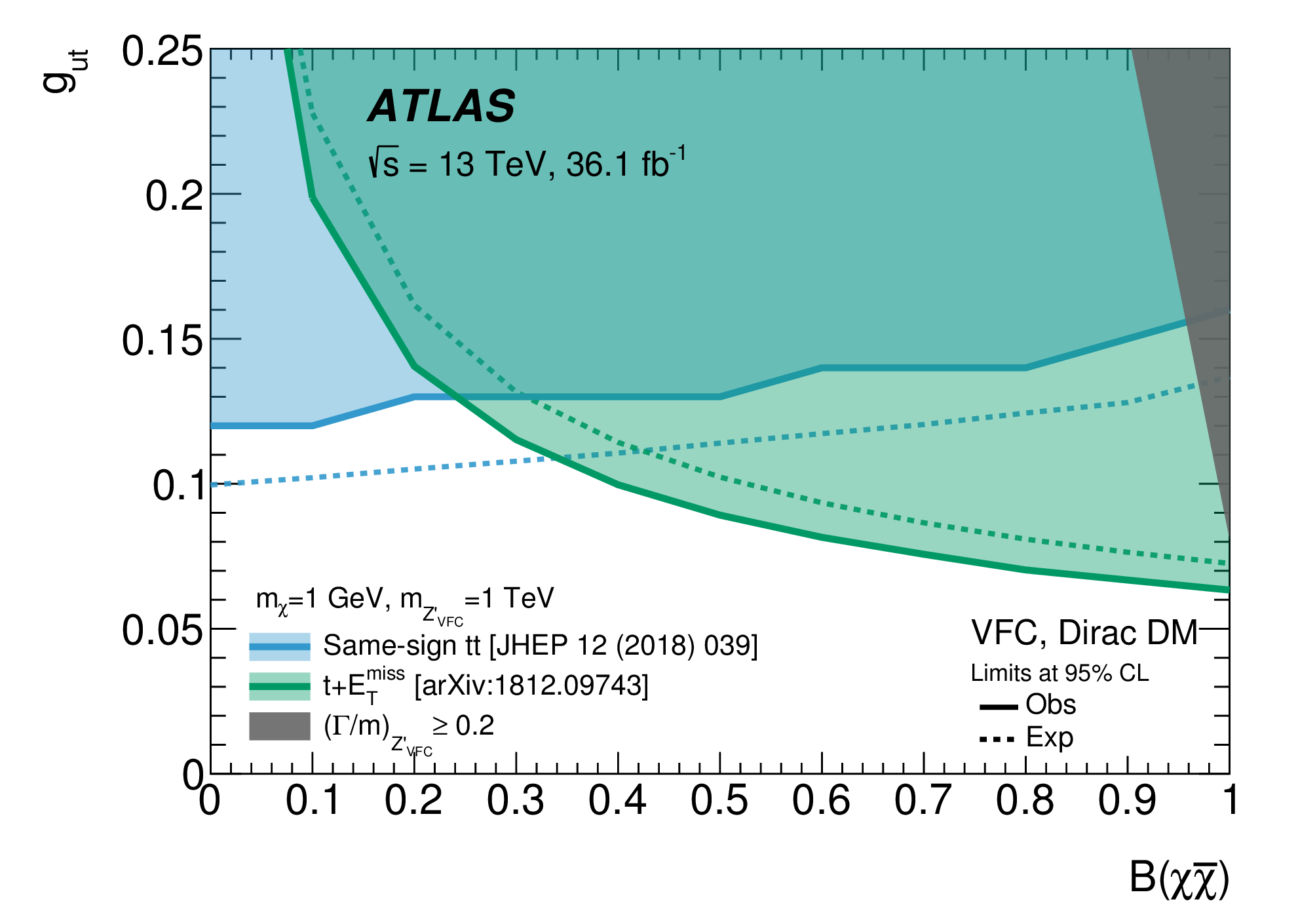}
\caption{Regions in the ($m_{Z'_{\mathrm{VFC}}}$,$g_{ut}$) (left) and the ($\mathcal{BR}(\chi \chi)$,$g_ut$) plane (right) of the VFC model excluded at 95\% CL by searches in the same-sign $t\bar{t}$ and \met+$t$ final states~\cite{ATLAS:2019wdu}.\label{fig:VFC_limits}}
\end{figure}   
\begin{figure}[h!]
\centering
\includegraphics[width=0.495\textwidth]{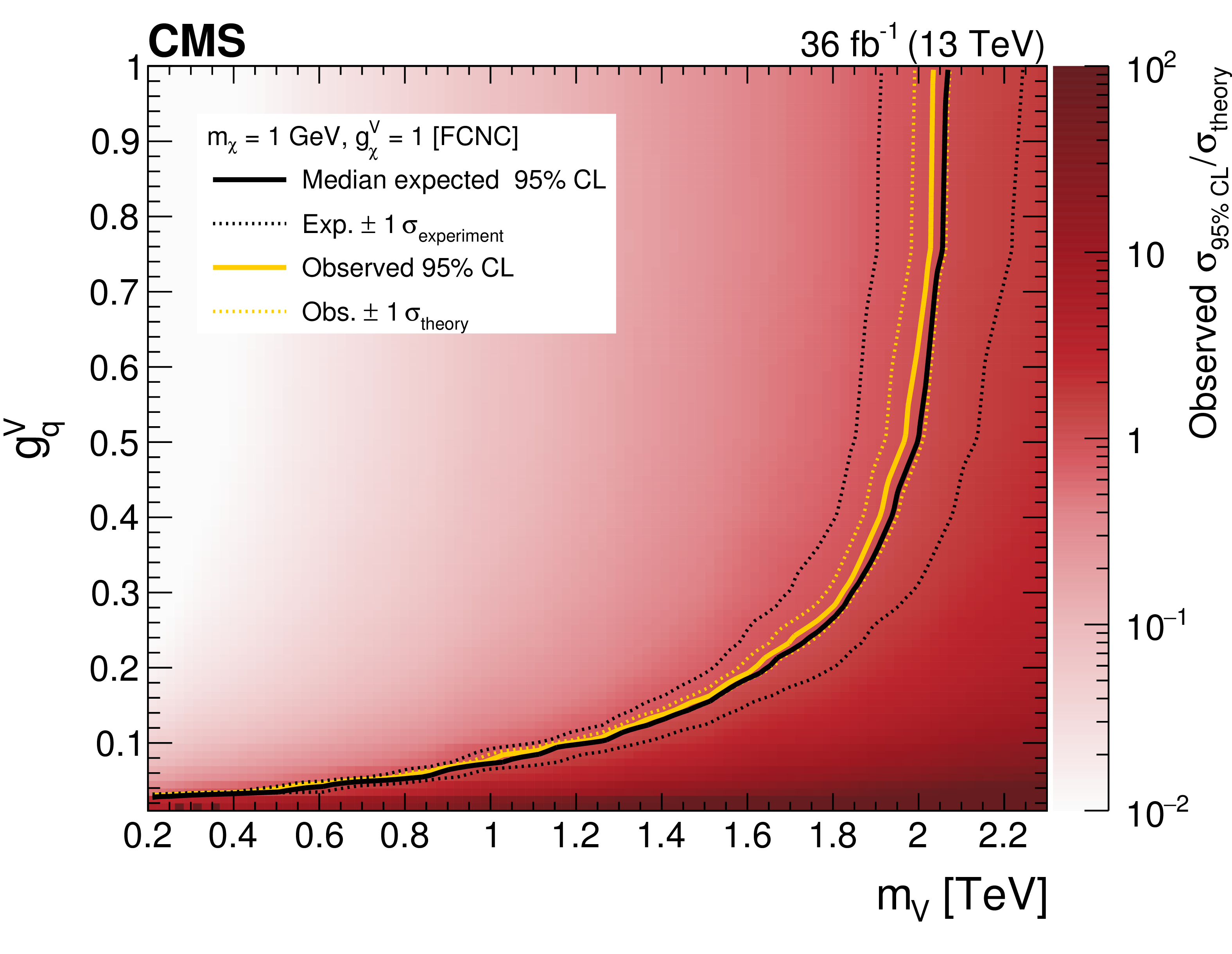}
\caption{Exclusion limits for the VFC model in the two-dimensional plane spanned by the mediator mass and the coupling between the mediator and quarks released by the CMS Collaboration~\cite{CMS:2018gbj}. The observed exclusion range is shown as yellow solid line, while the yellow dashed lines show the cases in which the predicted cross section is shifted by the assigned theoretical uncertainty. The expected exclusion range is indicated by a black solid line, the experimental uncertainties are shown in black dashed lines.\label{fig:VFC_limits_CMS}}
\end{figure}   

\subsection{Scalar and pseudoscalar mediators}

\subsubsection{Colour-neutral interaction}
\label{sec:SPS_results}

Simplified models with a colour-neutral scalar or pseudoscalar mediator have been constrained by searches targeting invisible mediator decays at the ATLAS and CMS experiments using data from $pp$ collisions at $\sqrt{s}=13$~TeV. The most recent  constraints from the CMS Collaboration based on \met+\ttbar events are shown in Figure~\ref{fig:SPS_limits_CMS}, while Figure~\ref{fig:SPS_limits} shows the most recent summary from the ATLAS Collaboration. 

Up to now, only \ttbar associated DM production has been probed by the CMS Collaboration using the full Run II dataset of 137~\ifb~\cite{CMS:2021eha}. The interpretation of this analysis in simplified models of scalar and pseudoscalar mediators is shown in Figure~\ref{fig:SPS_limits_CMS}. Assuming a mediator coupling of 1 to DM and SM particles, masses up to 400~\gev and 420~\gev can be excluded for scalar and pseudoscalar mediators, respectively. While the sensitivities of the 0- and 1-lepton channels are comparable, the sensitivity of the 2-lepton channel is significantly weaker. The sensitivity of this channel can be further enhanced by exploring information sensitive to the spin of the mediator which has not been done here. The exclusion limits for pseudoscalar mediators can be further extended up to 470~\gev by \met+jet searches~\cite{CMS:2021far}.

The results shown in Figure~\ref{fig:SPS_limits} are obtained from analyses targeting \met+$t\bar{t}$, \met+$tW$, \met+$tj$, \met+$b\bar{b}$, and \met+jet production using the full ATLAS Run 2 dataset of 139~fb$^{-1}$~\cite{ATLAS:DMSum}.
The sensitivity across most of the mediator mass region is dominated by a statistical combination of three searches for \met+$t\bar{t}$ production in the 0-, 1-, and 2-lepton channels (Section~\ref{sec:sig_inv_ttmet}).
In the scenario with a scalar mediator, the statistical combination of the \met+$t\bar{t}$ searches provides the strongest constraints across the probed mediator mass range, while pseudoscalar case, the dominant constraints for $m_{\phi/a}>300$~GeV are obtained from \met+jet searches.
Searches targeting the \met+$b\bar{b}$ signature provide significantly weaker constraints on this model. However, as explained in Section~\ref{sec:SPS_model}, in UV completions of the simplified model the couplings to up-type quarks can be suppressed compared to those to down-type quarks, making \met+$b\bar{b}$ searches a relevant complement to \met+$t\bar{t}$ searches.
Searches targeting DM production with a single top quark (\met+$tj$ and \met+$tW$, see Section~\ref{sec:SPS_model}) have a similar sensitivity as the individual searches for \met+$t\bar{t}$ production. They have not been included in the statistical combination as they are not orthogonal to the searches in the \met+\ttbar final states by construction.

\begin{figure}[p]
\centering
\includegraphics[width=0.495\textwidth]{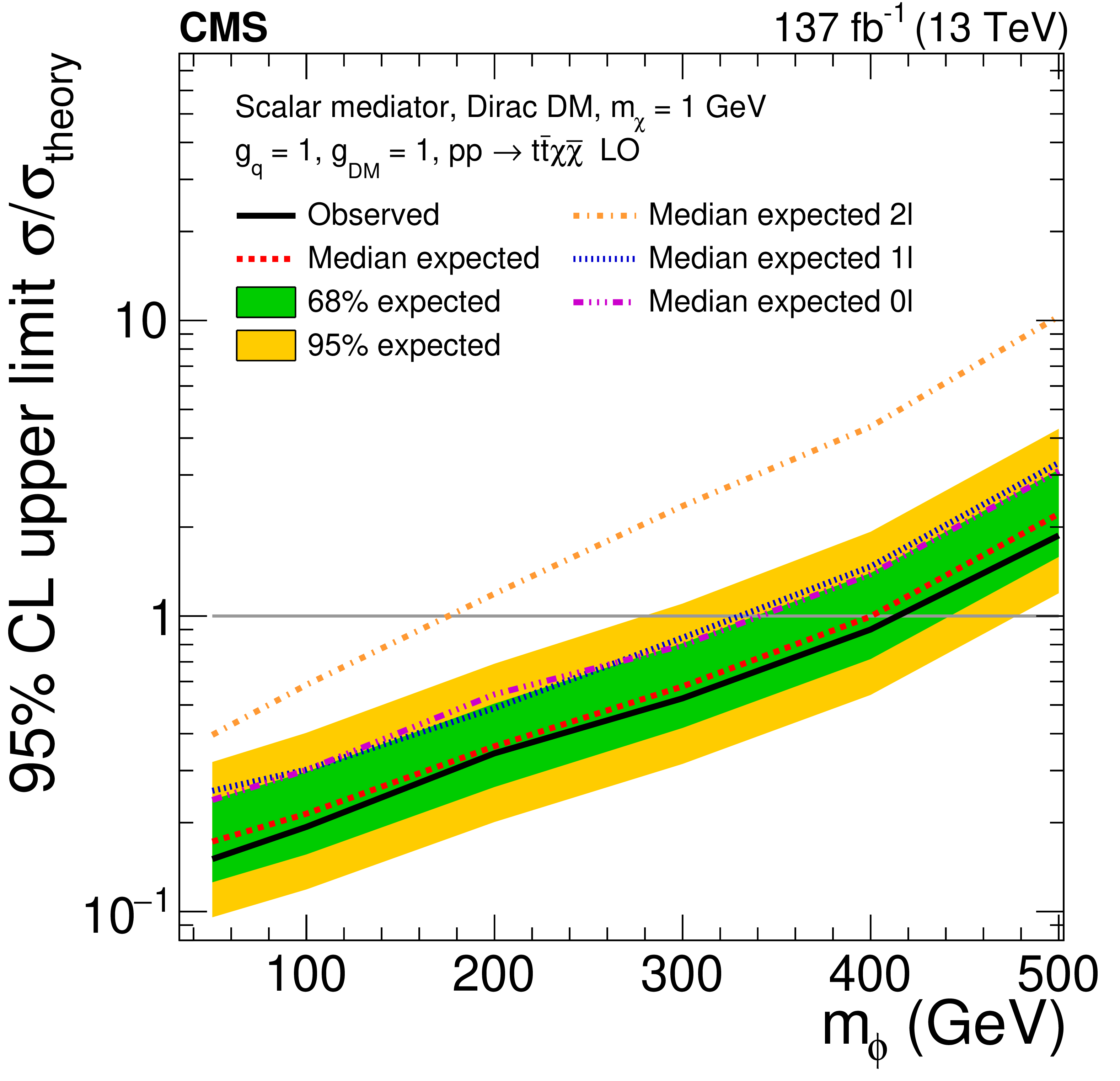}
\includegraphics[width=0.495\textwidth]{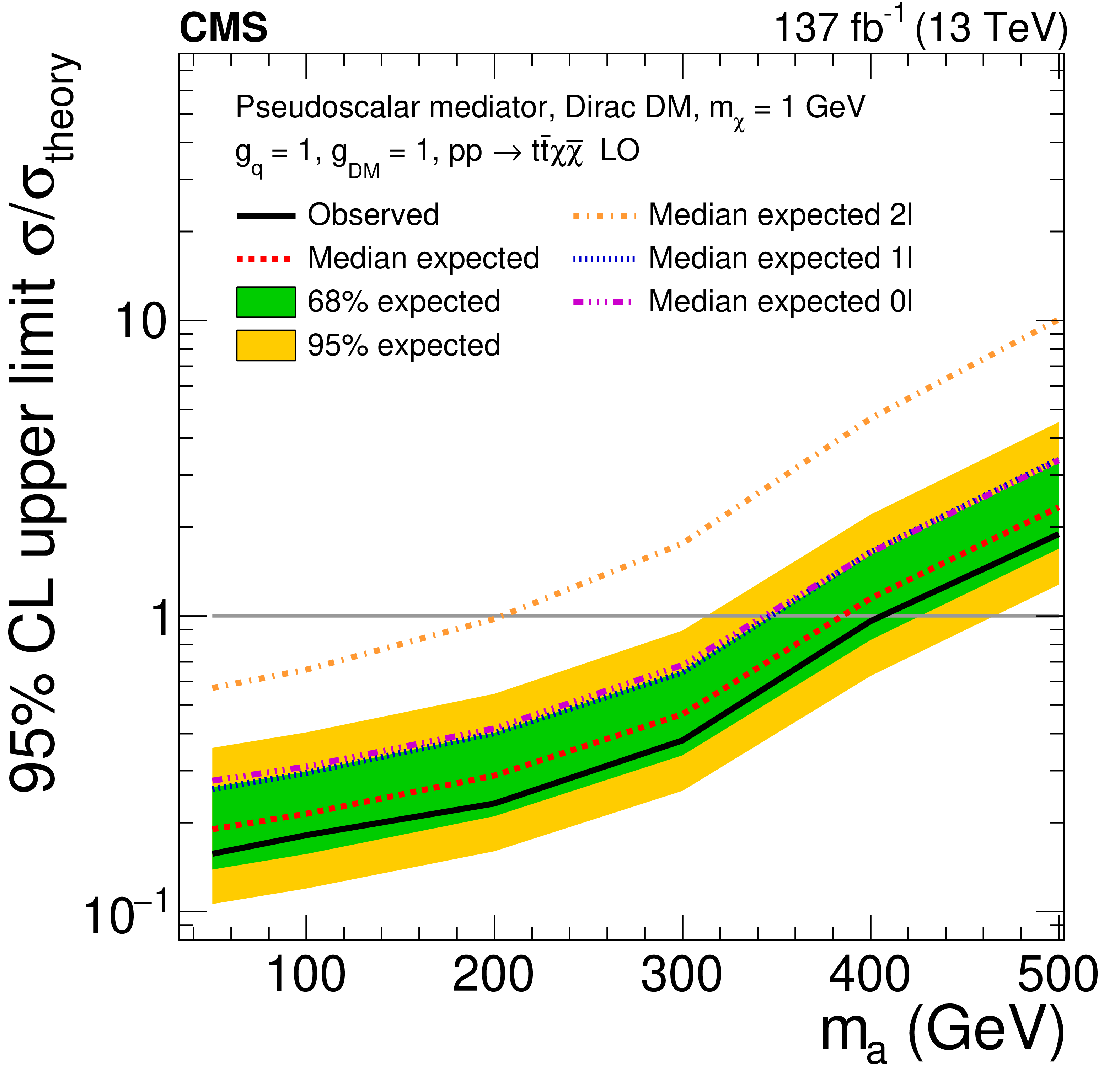}
\caption{Expected (dashed line) and observed (solid line) upper limits at the 95\% CL on the ratio of the excluded and predicted cross-section at leading-order for a DM particle with a mass of 1~\gev as a function of the mediator mass for a scalar (left) and pseudoscalar (right) mediator~\cite{CMS:2021eha}. The green and yellow bands represent the regions containing 68 and 95\%, respectively, of the distribution of limits expected under the background-only hypothesis. The mediator couplings are set to 1.\label{fig:SPS_limits_CMS}}
\end{figure}   

\begin{figure}[p]
\centering
\includegraphics[width=0.495\textwidth]{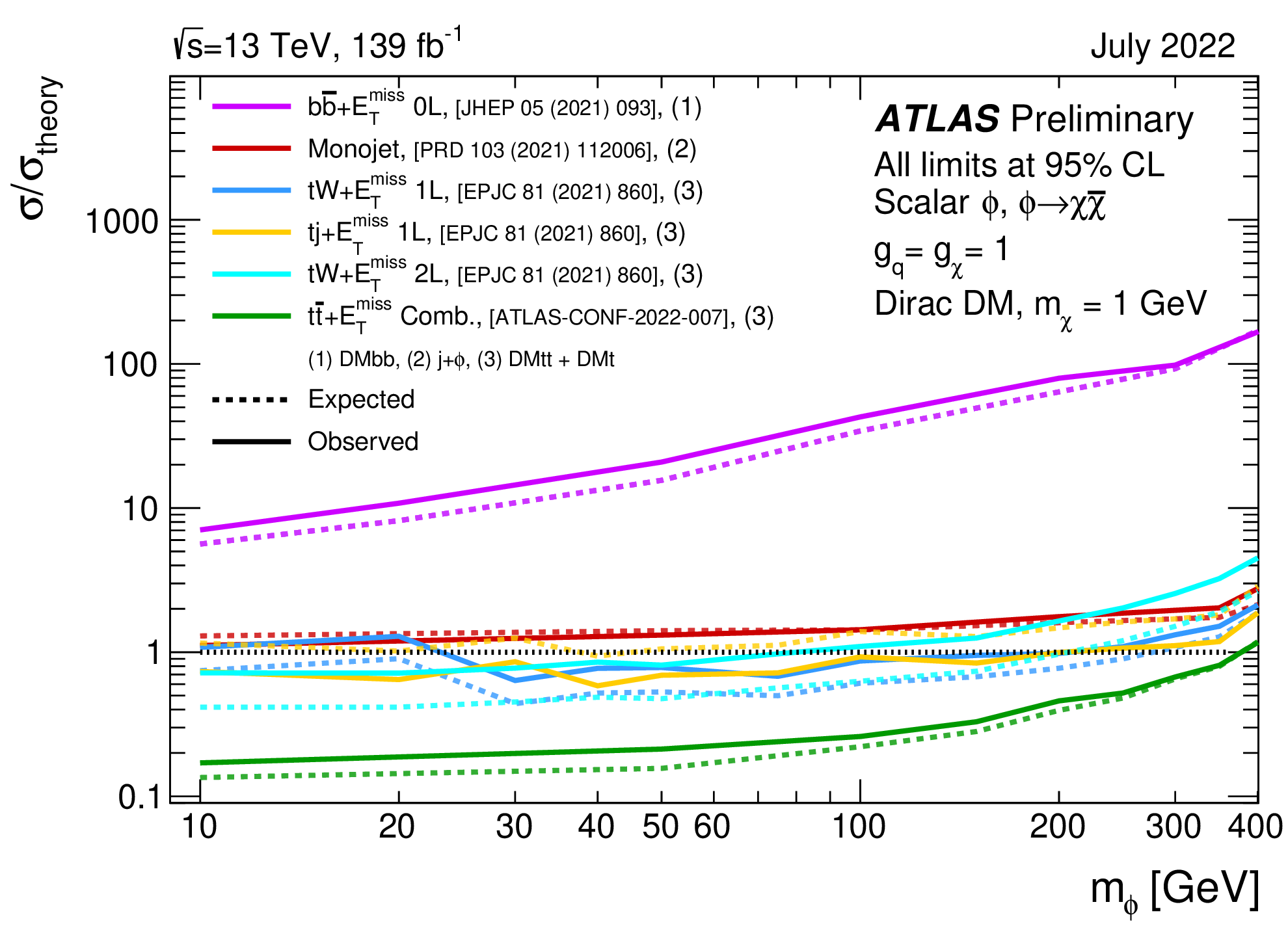}
\includegraphics[width=0.495\textwidth]{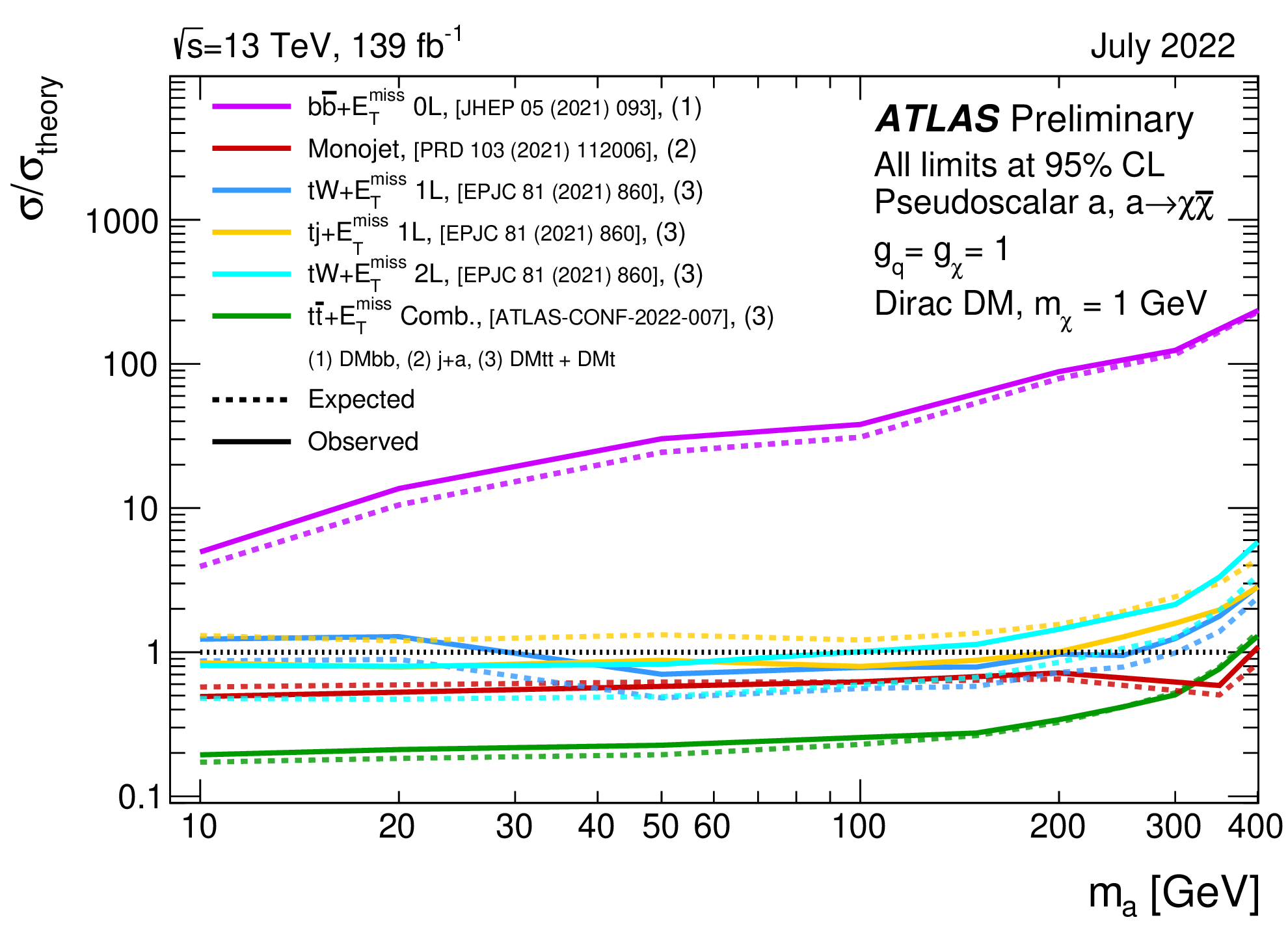}
\caption{Upper limits at 95\% CL on the production of a scalar $\phi$ (left) and pseudoscalar $a$ (right) mediator as a function of the mediator mass~\cite{ATLAS:DMSum}. The limits are expressed in terms of the ratio of the excluded cross-section and the cross-section calculated for a coupling assumption of $g=g_q=g_{\chi}=1.0$. The latter was calculated at NLO for the \met+\ttbar signatures and at LO for the \met+$tW$/$tj$ and \met+$j$ signatures.
\label{fig:SPS_limits}}
\end{figure}   

If $m_{\phi/a} > 2\cdot m_{t}$, searches targeting visible mediator decays to top quarks are also sensitive to the production of scalar or pseudoscalar mediators. Two different modes can contribute: gluon-induced mediator production and production of a mediator in association with \ttbar.
Searches targeting both modes have been performed, as discussed in Sections~\ref{sec:sig_vis_ttbar} and~\ref{sec:sig_vis_4top}, respectively.
However, only the results of a search for four-top production conducted by the CMS Collaboration have been interpreted in the context of simplified models with a scalar or pseudoscalar mediator.
The results are shown in Figure~\ref{fig:SPS_limits_4top} as upper limits on the cross-section of associated production of the mediator with top quarks times the branching ratio of the mediator decay to \ttbar. Masses between 350~GeV and 450 (510) GeV for a scalar (pseudoscalar) mediator are excluded.

\begin{figure}[H]
\centering
\includegraphics[width=0.495\textwidth]{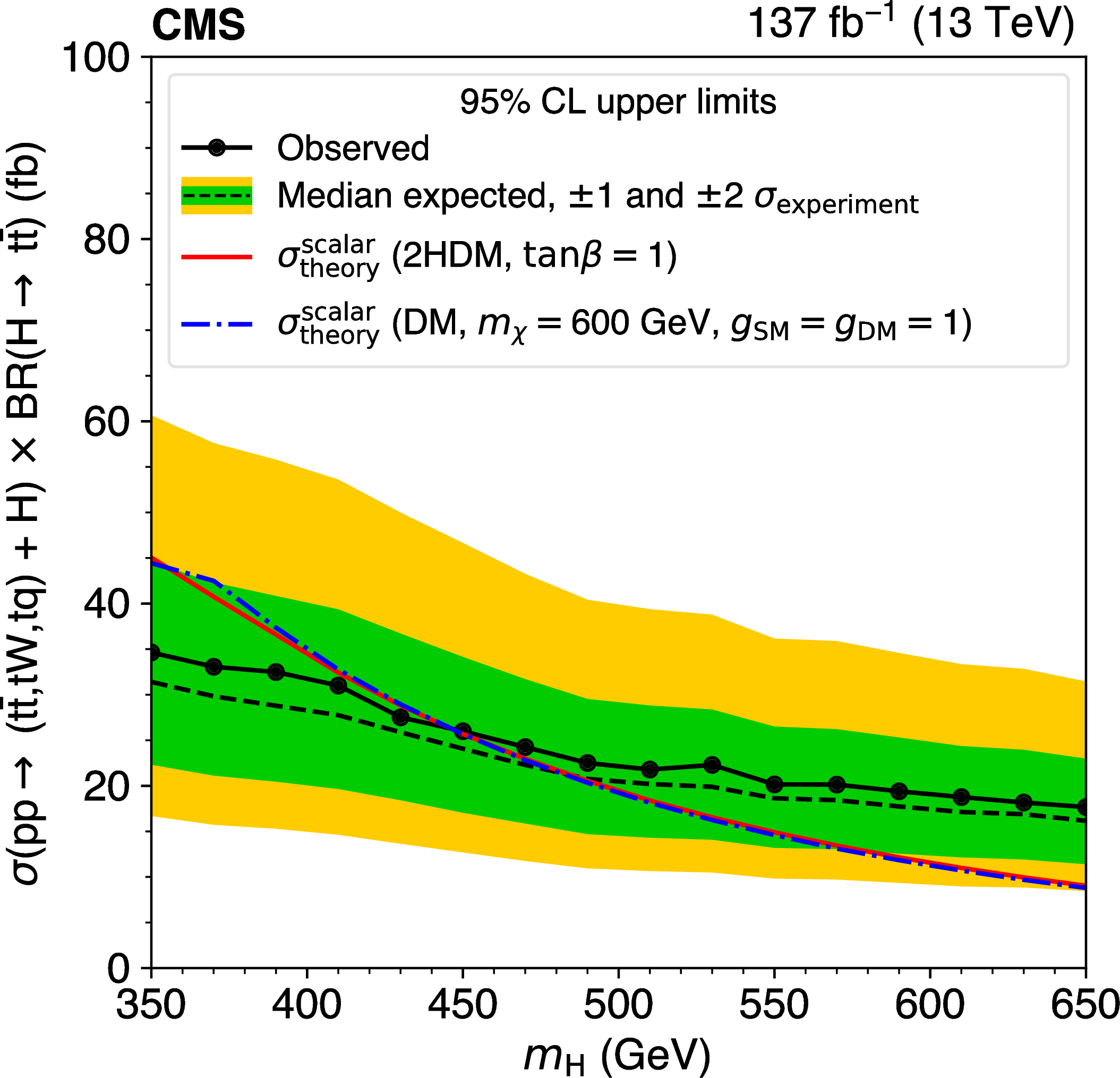}
\includegraphics[width=0.495\textwidth]{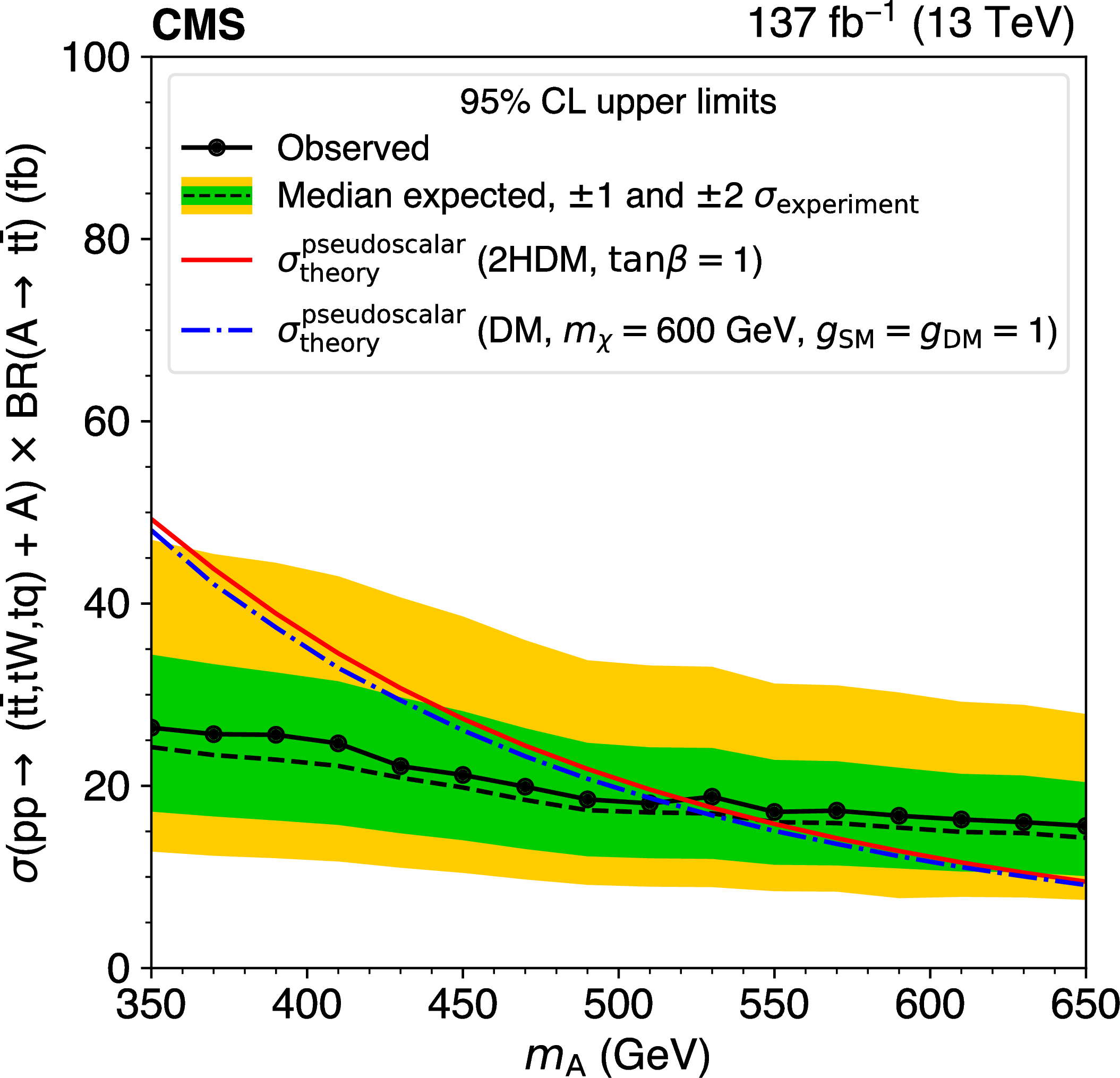}
\caption{Upper limits at 95\% CL on the production of a scalar (left, called $H$ here instead of $\phi$) and pseudoscalar (right, called $A$ here instead of $a$) mediator as a function of the mediator mass~\cite{ATLAS:DMSum}. The limits are expressed in terms of an upper limit on the production cross-section times the branching ratio of the mediator to \ttbar and compared to the cross-section calculated at LO for a coupling assumption of $g=g_q=g_{\chi}=1.0$ (here denoted as: $g_{\mathrm{SM}}=g_{\mathrm{DM}}=1.0$).
\label{fig:SPS_limits_4top}}
\end{figure} 

It should be noted that the re-interpretation of the results from searches targeting gluon-induced mediator production is significantly more involved than for the case of associated production due to the presence of strong signal-background interference (Section~\ref{sec:sig_vis_ttbar}). The resulting interference patterns are highly model-dependent which means that a re-interpretation in the context of a different model requires the generation of the model-specific interference pattern and a subsequent re-running of the full profile likelihood fit for these model-specific interference patterns.

\subsubsection{Colour-charged interaction}
\label{sec:SCC_results}

Models in which the colour-charged mediator decays to a top quark and a DM particle are constrained by the searches in \met+$t$ final states discussed in Section~\ref{sec:sig_inv_mett}. 
Mediator masses up to 5~TeV can be excluded by the ATLAS Collaboration for coupling strength values $\lambda_t=0.4$ and $g_{ds}=0.6$ assuming a DM mass $m_{\chi}=10$~GeV~\cite{ATLAS-CONF-2022-036}. 

Results with a mixed scalar and pseudoscalar coupling to both SM quarks as well as DM and top quarks are provided by CMS Collaboration~\cite{CMS:2018gbj}. Assuming a coupling of 0.1 to SM quarks and of 0.2 to DM and top quarks, mediators with masses up to 3.3~\tev can be excluded for a dark matter mass of 100~\gev.

\subsection{Extended Higgs sectors}

\subsubsection{2HDM with a pseudoscalar mediator}
\label{sec:2HDMa_results}

Constraints on the 2HDM+$a$ are derived from a variety of searches targeting different production and decay modes of the mediator and the additional Higgs bosons. 
The most comprehensive summary of constraints has been released by the ATLAS Collaboration~\cite{ATLAS:DMSum}. These summary plots are based results obtained on the partial or full Run 2 datasets. Not all of the latest searches on the full Run 2 dataset have been re-interpreted in the context of the 2HDM+$a$. Updated summary plots will be released in the near future.

The constraints are evaluated as a function of the free parameters of the model described in Section~\ref{sec:2HDMa_model}. 
Two representative parameter scans in the ($m_a$,$m_{A}$) and the ($m_a$,$\tan\beta$) plane highlighting the interplay of signatures involving top quarks with other types of signatures are shown in Figure~\ref{fig:2HDMa_limits}. The constraints for other benchmark scans can be found in Ref.~\cite{ATLAS:DMSum}.

\begin{figure}[h!]
\centering 
\includegraphics[width=0.495\textwidth]{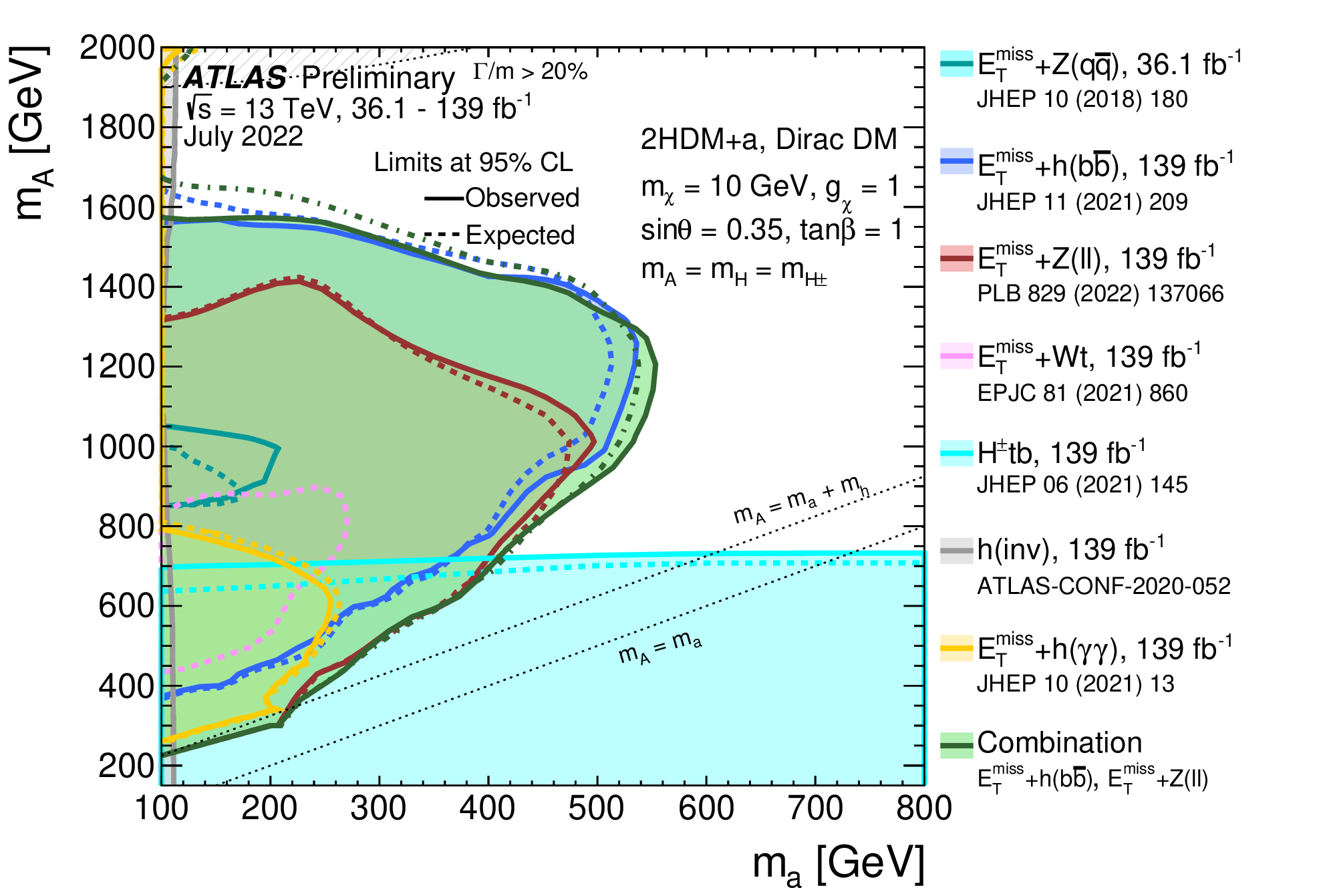}
\includegraphics[width=0.495\textwidth]{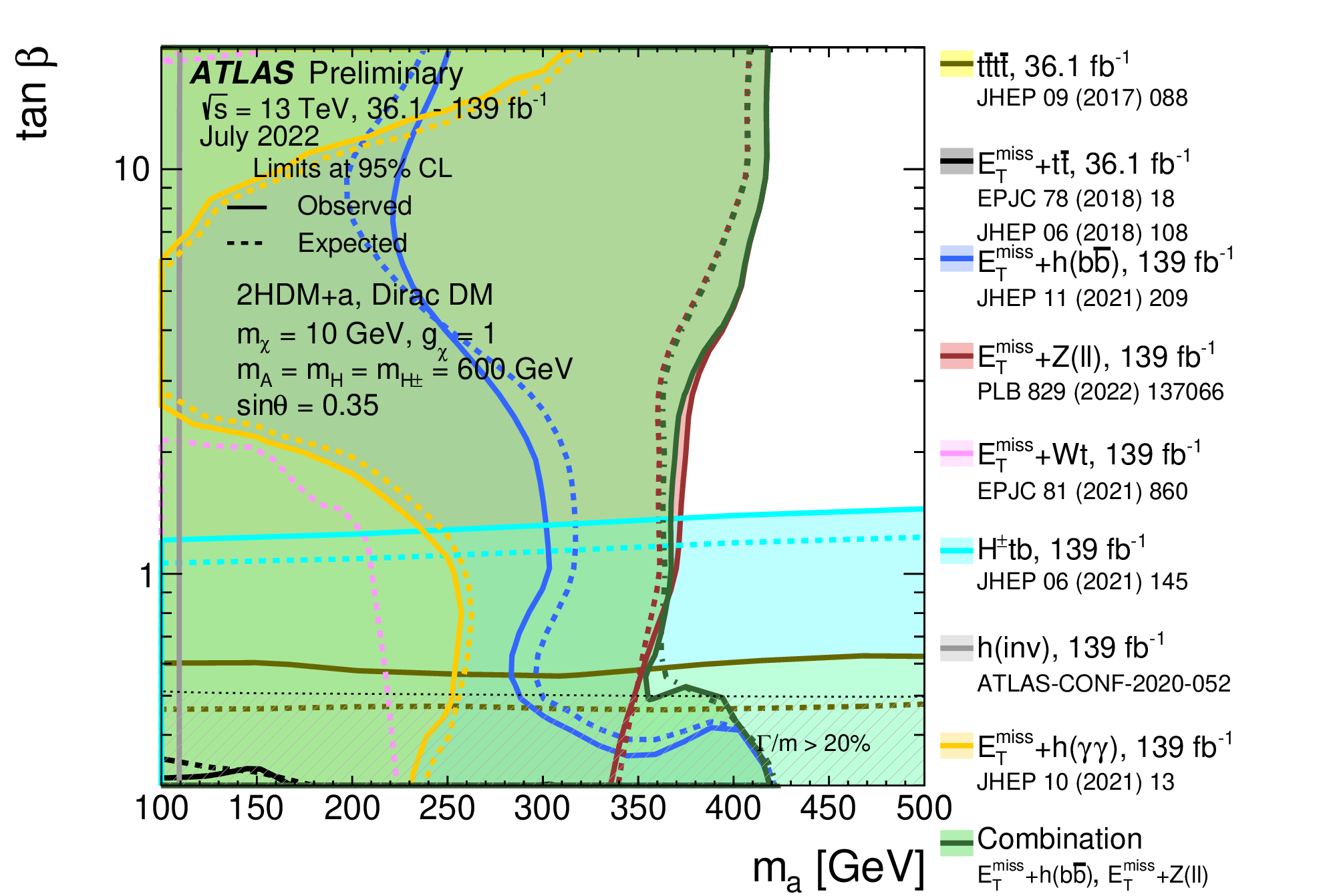}
\caption{Regions in the 2HDM+$a$ parameter space excluded at 95\% CL by several individual searches targeting different signatures and a statistical combination of \met+$Z(\ell\ell)$ and \met+$h(b\bar{b})$ searches. The results are shown in the ($m_a$,$m_{A}$) plane (left) and the ($m_a$,$\tan\beta$) plane (right). In the former case, $\tan\beta=1$, while in the latter case, $m_A = 600$~GeV. In both cases, the conditions $\sin\theta=0.35$ and $m_A = m_H = m_{H^{\pm}}$ are imposed. All results are based on either the full 139~fb$^{1}$ of $pp$ collision data at $\sqrt{s}=13$~TeV or a subset of that dataset amounting to 36~fb$^{1}$~\cite{ATLAS:DMSum}.\label{fig:2HDMa_limits}}
\end{figure}   

The sensitivity in the ($m_a$,$m_{A}$) plane for $\tan\beta=1$, $\sin\theta=0.35$, and $m_A = m_H = m_{H^{\pm}}$ is largely dominated by searches targeting the production of an invisibly decaying mediator with a Higgs or $Z$ boson, leading to \met+$h$ and \met+$Z$ signatures, directly. These processes are dominated by diagrams involving the resonant production of a neutral Higgs bosons $H$ or $A$ that decays to $ah$ or $aZ$, respectively. The sensitivity from searches for \met+$tW$ production, which can also proceed resonantly via a charged Higgs boson (Section~\ref{sec:2HDMa_model}) is sub-dominant in this parameter region.

Constraints that are largely complementary to those from \met+$X$ searches are obtained from a search targeting resonant associated production of a charged Higgs boson $H^{\pm}$ with a top-bottom quark pair ($tbH^{\pm}$) with subsequent decay to a top-bottom quark pair $tb$. These constraints exhibit only a weak dependence on the mediator mass $m_a$ as this signature does not involve production of a mediator at leading order and is hence only indirectly dependent on the mediator mass via its effect on the branching ratio to $tb$ compared to those for other decays, such as $H^{\pm}\rightarrow aW^{\pm},AW^{\pm},HW^{\pm}$.

Searches targeting resonant production of the neutral Higgs bosons $A/H$, either via gluon fusion or $t\bar{t}$ associated production, and their decay to $t\bar{t}$, leading to $t\bar{t}$ and $t\bar{t}t\bar{t}$ final states, respectively, are expected to also provide complementary constraints to those from \met+$X$ searches in this parameter region, given that the choice $\tan\beta=1$ favours the coupling of those Higgs bosons to top quarks.
No constraints from $A/H(t\bar{t})$ have been derived for the 2HDM+$a$ yet due to the presence of strong, model-dependent interference effects that make a straightforward re-interpretation of these searches in the context of other benchmark models difficult, as explained in Section~\ref{sec:SPS_results}.
A search targeting $t\bar{t}A/H(t\bar{t})$ production has been used to constrain the 2HDM+$a$ parameter space (see below). It is based on 36~fb$^{-1}$ of LHC Run 2 data and not sensitive at $\tan\beta=1$, as shown in Figure~\ref{fig:2HDMa_limits} (right plot). The results of a search for $t\bar{t}A/H(t\bar{t})$ production in multi-lepton final states using 139~fb$^{-1}$ of LHC Run 2 data indicate that $A/H$ masses up to 700~GeV could be excluded in the 2HDM+$a$ for the parameter region with $\tan\beta$ under consideration here~\cite{ATLAS:2022rws}.

In the ($m_a$,$\tan\beta$) plane with $m_{A}=m_{H}=m_{H^{\pm}}=600$~GeV (right plot in Figure~\ref{fig:2HDMa_limits}), the sensitivity is again dominated by the statistical combination of the \met+$h(b\bar{t})$ and \met+$Z(\ell\ell)$ searches and the search for $tbH^{\pm}(tb)$ production, which provide complementary constraints in this region of parameter space. Low values of $\tan\beta$ are fully excluded by the search for charged Higgs bosons decaying to $tb$. The constraints from the search targeting $t\bar{t}t\bar{t}$ production on 36~fb$^{-1}$ of LHC Run 2 data are also shown. While they are notably weaker than the constraints from the charged-Higgs-boson search, which relies on the full Run 2 dataset amounting to 139$^{-1}$, the results from the search for $t\bar{t}A/H(t\bar{t})$ on 139~fb$^{-1}$ of LHC Run 2 data~\cite{ATLAS:2022rws}
(Section~\ref{sec:sig_vis_4top}) indicate that this final state may provide a comparable exclusion power as the charged-Higgs-boson search if re-interpreted in the context of this model.

Searches for \met+$t\bar{t}$ production, which dominate the sensitivity to the simplified model with a colour-neutral scalar or pseudoscalar mediator (Section~\ref{sec:SPS_results}), only weakly constrain the benchmark scenarios~\cite{LHCDarkMatterWorkingGroup:2018ufk,ATLAS:2HDMa_2021} probed at the LHC. It should, however, be noted that the \met+$t\bar{t}$ constraints shown in Figure~\ref{fig:2HDMa_limits} are based on only 36~fb$^{-1}$ of LHC Run 2 data and the sensitivity is mainly limited by low event rates. Hence significantly stronger constraints are expected from a re-interpretation of searches using the full 139~fb$^{-1}$ of LHC Run 2 data~\cite{ATLAS:METplustt}. The sensitivity of the \met+$t\bar{t}$ final state is expected to become comparable to that of searches in the \met+$h$ and \met+$Z$ final states for an integrated luminosity of 300~fb$^{-1}$, expected to be available after the end of LHC Run 3 (2022--2025)~\cite{Bauer:2017ota}.
In this context, it should be noted that the cross-section for \met+$t\bar{t}$ production is suppressed by $\sin\theta^2$, making this process more sensitive for large values of $\sin\theta$~\cite{Bauer:2017ota}. Furthermore, for $m_a > 2\cdot m_t$, visible mediator decays to $t\bar{t}$ are possible, reducing the invisible branching ratio $a \rightarrow \chi\chi$ and hence the sensitivity of the \met+$t\bar{t}$ searches~\cite{Bauer:2017ota}.

\subsection{Scalar DE EFT model}
\label{sec:DE_results}

Searches in the \met+$t\bar{t}$ final state have been used to constrain the $\mathcal{L}_1$ operator in the EFT model of scalar DE (Section~\ref{sec:DE_model}).
Results from three independent analyses, each targeting a different $t\bar{t}$ decay mode (0-, 1-, 2-lepton channels) have been used. No statistical combination was performed. Instead, the constraint from the analysis yielding the smallest CL$_{\mathrm{s}}$ value for a given signal hypothesis was re-interpreted in the EFT model of DE.
The strongest constraints arise from searches in the 0- and 1-lepton channels, with both contributing roughly equally.

The constraints are derived as a function of the effective coupling $g_*$ associated with the UV completion of the EFT model and the effective mass scale $M_1$. 
It is assumed that the EFT is valid for momentum transfers $Q_\textrm{tr} < g_* M$~\cite{ATLAS:2019wdu}.
For events failing this requirement, a conservative approach to correct the final limits based on the fraction of valid events, referred to as iterative rescaling~\cite{Abercrombie:2015wmb}, is applied.

The regions excluded at 95\% CL are shown in Figure~\ref{fig:DE_limits}.
Mass scales $<200$~GeV are excluded for $g_*>\pi^2$.
The sensitivity of the \met+$t\bar{t}$ signature to softer effective couplings $g_*$ is limited
by the EFT criterion as $t\bar{t}$ pair production typically involves large momentum transfers.

\begin{figure}[H]
\centering
\includegraphics[width=0.5\textwidth]{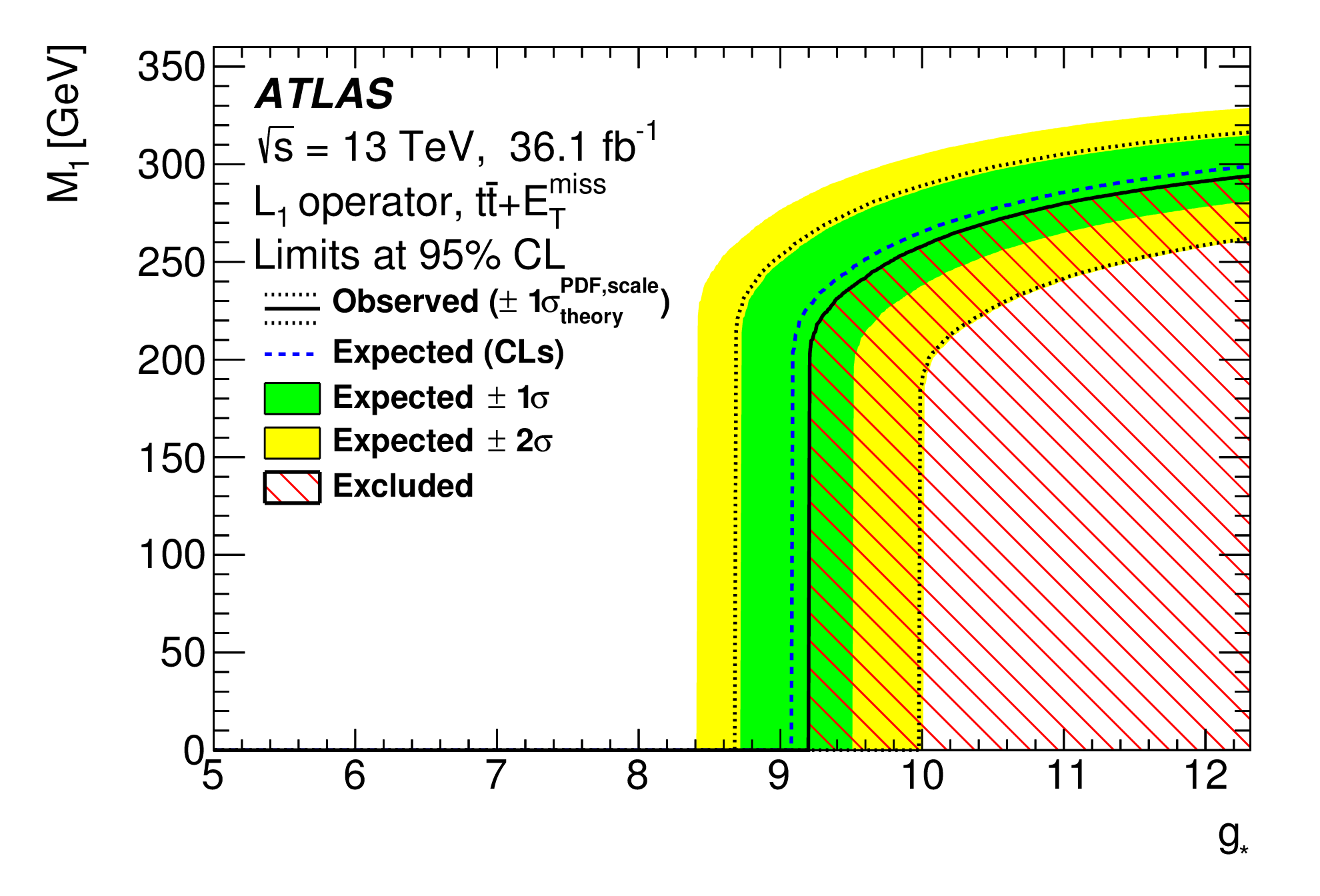}
\caption{Regions in the plane of the effective coupling $g_*$ associated with the UV completion of the EFT model and the effective mass scale $M_1$ for the $\mathcal{L_1}$ operator excluded at 95\% CL by searches in the \met+$t\bar{t}$ final state~\cite{ATLAS:2019wdu}.\label{fig:DE_limits}}
\end{figure}   

\section{Discussion}
\label{sec:discussion}

A variety of searches targeting top-quark production in association with DM or via visible decays of mediator particles have been conducted by the ATLAS and CMS Collaborations. No significant deviation from the SM prediction has been observed. Therefore the results are used to constrain DM in a variety of simplified models as well as scalar DE described in an EFT model.
Signatures involving top quarks often provide sensitivity in parameter regions not covered by other DM searches, underlining their importance as sensitive probes of DM at colliders. They provide a particularly relevant probe of models involving new particles with Yukawa-like interactions, which imply preferred couplings to top quarks.

It should be noted that many of the results and summary plots presented in this review are preliminary as various searches on the full LHC Run 2 collision data still on-going. Furthermore, not all of the existing results have been interpreted in relevant benchmark models. Further results of DM searches with top quarks are expected to be released by both collaborations in the near future.

\section{Outlook}
\label{sec:outlook}

\subsection{LHC Run 3}
\label{sec:LHCRun3}

The non-observation of WIMP DM at the LHC and various direct detection experiments to date has prompted the particle physics community to place a stronger focus on models and searches for non-WIMP DM as well as uncovered DM signatures at the LHC that can be probed during LHC Run 3 (2022-2025) and/or via re-interpretations of existing searches on LHC Run 2 data. A few notable examples involving signatures with top quarks are given in the following.

\subsubsection{ALPs}
\label{sec:ALPs}

Axions and axion-like particles (ALPs)~\cite{ALPs1,ALPs2} have received increasing attention in recent years. A novel strategy to search for ALPs and, more generally, pseudo-Nambu-Goldstone bosons (pNGB) at the LHC has been proposed in Ref.~\cite{Gavela:2019cmq}, focusing on non-resonant searches that would be sensitive to ALPs produced as an off-shell $s$-channel mediator. It is motivated by the fact that the pNGB nature of the ALPs implies that their couplings to the SM are dominantly derivative, which leads to a cross-section enhancement for non-resonant ALPs production at centre-of-mass energies $\hat{s}>>m_a$, where $m_a$ denotes the mass of the ALP.
The focus of recent studies has been on constraining the ALP-boson ($W$, $Z$, $h$, $g$, $\gamma$) coupling via non-resonant $ZZ$, $\gamma\gamma$, and $gg$~\cite{Gavela:2019cmq}, non-resonant $ZZ$ and $Zh$~\cite{CMS:2021xor}, and non-resonant $WW$, $Z\gamma$~\cite{Carra:2021ycg} production.
The ALPs-fermion coupling can be predominantly probed via non-resonant \ttbar production (illustrated by the left diagram in Figure~\ref{fig:ALPs}) due to the Yukawa-like structure of the ALP-fermion couplings. No public results exist to date but studies are on-going.

\begin{figure}[H]
\centering 
\includegraphics[width=0.35\textwidth]{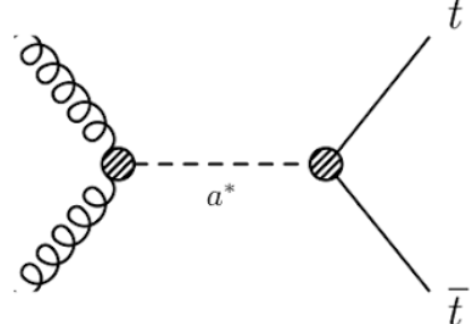}
\includegraphics[width=0.425\textwidth,height=3.5cm]{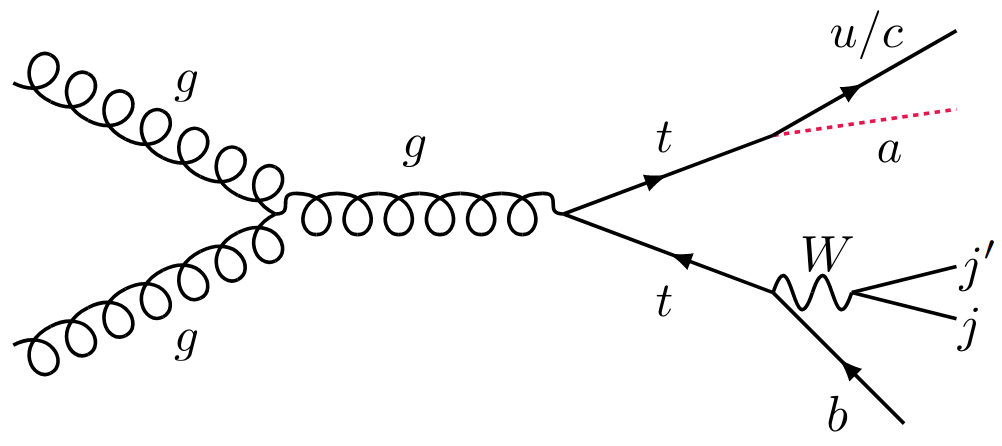}
\caption{Schematic representation of non-resonant \ttbar production via an off-shell $s$-channel ALP (left,~\cite{ALPS_NR_Feyn}) and SM \ttbar production with subsequent decay of one of the top quarks to an up-type quark and a long-lived ALP (right,~\cite{Carmona:2022jid}).\label{fig:ALPs}}
\end{figure}   

The ALPs-fermion coupling can also be probed in \met+\ttbar final states. These are sensitive to \ttbar-associated production of a single ALP with couplings to quarks derived from couplings to the bosonic sector and proportional to the fermion mass~\cite{Brivio:2017ije}. It should be noted that the \met distribution predicted for this signal process is softer on average than that predicted by e.g. stop production in supersymmetric models, emphasising the importance of keeping the \met threshold low in future searches.

Novel detector signatures involving exotic top quark decays are predicted in models with flavour-violating ALPs~\cite{Carmona:2022jid}, which are motivated by $t$-channel dark sector models~\cite{Renner:2018fhh} or Frogatt-Nielsen models of flavour~\cite{FVALPs}.
These models predict flavour-violating decays of the top quark to an up-type quark and an ALP, with the ALP decaying predominantly to hadrons, either promptly or with a long lifetime. Precision measurements of single-top-quark production can constrain the parameter space of such models for prompt ALPs decays to jets and detector-stable ALPs. Displaced detector signatures are predicted for non-prompt ALPs decays within the detector volume. A novel search has been proposed~\cite{Carmona:2022jid} focusing on exotic top-quark decays from SM \ttbar production (right diagram in Figure~\ref{fig:ALPs}), where one of the top quarks decays into an up-type quark and an ALP, which in turn decays into a displaced narrow jet within the calorimeter volume.
This and other signatures involving long-lived particles (LLP) in top-quark decays have not yet been probed in dedicated searches at the LHC.
They remain an exciting prospect for the analysis of LHC Run 3 data within the currently fast-growing field of LLPs searches at the LHC, a field that benefits in particular from novel trigger and reconstruction algorithms deployed by the ATLAS and CMS experiments for Run 3 data taking.

\subsubsection{Composite pseudo Nambu-Goldstone Bosons}
\label{sec:pNGB}

Signatures with top quarks can also be used to probe still viable WIMP models in which WIMP DM is made up of composite pNGBs~\cite{Haisch:2021ugv}. In these models, both the SM Higgs boson and DM emerge from a TeV-scale strongly-coupled sector as pNGBs and the SM-DM interaction is provided by higher-dimensional derivative couplings with the Higgs fields, which leads to a strong suppression of the DM scattering rates against SM particles. Thus, these models evade the strong constraints from direct detection experiments, making collider searches particularly relevant. The pNGB DM contains additional interactions with the SM sector, besides the derivative Higgs portal, with preferential couplings to third-generation fermions being well-motivated~\cite{Haisch:2021ugv}.
If couplings to top quarks are preferred over couplings to bottom quarks, e.g. in the case of Yukawa-type couplings, pNGB models can be probed at the LHC via associated production of pNGB DM with \ttbar or a single top quark, i.e. in \met+\ttbar or \met+$t$+X final states. Two possible production modes of pNGB leading to \met+$tW$ final states via the Higgs portal and direct DM-top interactions are shown in Figure~\ref{fig:pNGBs}.
Searches in these final states are complementary to searches for invisible Higgs boson decays in vector-boson fusion (VBF) production as they are sensitive to pNGB interactions with fermions not accessible via the latter. Re-interpretations of existing \met+\ttbar and \met+$tW$ searches as well as possible optimisations of future searches for pNGB production could be interesting to explore during LHC Run 3.

\begin{figure}[H]
\centering
\includegraphics[width=0.35\textwidth]{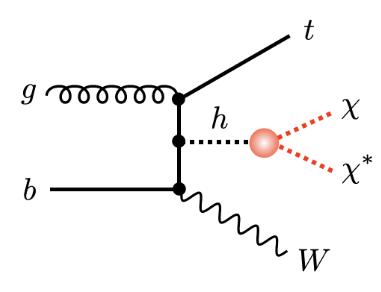}
\includegraphics[width=0.35\textwidth]{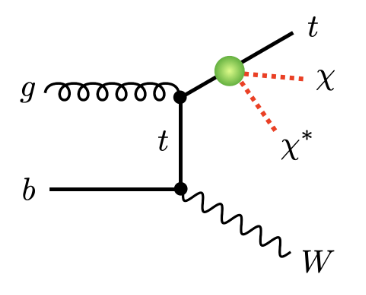}
\caption{Schematic representation of \met+$tW$ production via DM-Higgs operators (left) and DM-top operators in an EFT of composite pNGBs~\cite{Haisch:2021ugv}.\label{fig:pNGBs}}
\end{figure}   

\subsubsection{Dark Mesons}
\label{sec:darkMesons}

Final states with multiple top quarks are predicted in models with a strongly coupled dark sector consisting of composite particles that carry electroweak but no colour charges~\cite{Kribs:2018ilo}. These models not only address the hierarchy problem but can also provide a DM candidate in the form of a composite meson whose decays are suppressed via an automatic accidental symmetry. 

The most promising target for collider searches is the dark meson sector, consisting of dark vector mesons $\rho_D$ and dark pions $\pi_D$~\cite{Kribs:2018ilo}.
Signatures with multiple top or bottom quarks are predicted if a pair of dark pions with gauge-phobic couplings to the SM is produced from the decay of a resonantly produced $\rho_D$ ($pp\rightarrow \rho_D \rightarrow \pi_D\pi_D$). The dark pions then decay predominantly into third-generation fermions, with decays to \ttbar ($tb$) dominating the branching fraction for $\pi_D^0$ ($\pi_D^{\pm}$) if the pion mass is above the \ttbar ($tb$) production threshold. Depending on the charge of the intermediate $\rho_D$, different final states involving third-generation quarks are possible: $b\bar{b}t\bar{b}$, $t\bar{t}b\bar{b}$, $t\bar{t}t\bar{b}$. 

Existing searches in multi-top final states only weakly constrain the parameter space of these models~\cite{Kribs:2018ilo}. This is due to the fact that small masses of the $\rho_D$ and $\pi_D$ are still viable, which means that the SM fermions in the final state tend to be rather soft. In searches at $\sqrt{s}=13$~TeV, in particular, higher thresholds are imposed on the energy/momenta of the final-state objects or their vector sum. In order to probe dark pions, or more generically strongly-coupled like models, dedicated searches targeting final states with a high multiplicity of low-momentum objects compatible with the decays of one or several low-momentum top quarks are needed.

\subsection{HL-LHC and HE-LHC}
\label{sec:out_HLLHC}

The physics potential for DM searches involving top quarks during the high-luminosity phase of the LHC (HL-LHC, starting 2028) and the perspectives for a possible future high-energy LHC (HE-LHC) have been studied in the context of a 2019 CERN Yellow Report~\cite{HLLHC_YR}.
The final HL-LHC dataset is expected to amount to an integrated luminosity of 3000~\ifb at a centre-of-mass energy $\sqrt{s}=14$~TeV. The HE-LHC scenario relies on the assumption of a possible further upgrade of the LHC to a 27~TeV $pp$ collider with a final integrated luminosity of 15,000~\ifb.

Sensitivity studies have been performed for the \met+\ttbar, \met+$tW$, \met+$t$, \ttbar, and $\ttbar\ttbar$ signatures within various benchmark models, including simplified models with a scalar or pseudoscalar mediator (Section~\ref{sec:SPS_model}), simplified models with a vector mediator with a flavour-changing coupling to the top and up quark (Section~\ref{sec:VFC_model}), and the 2HDM+$a$ (Section~\ref{sec:2HDMa_model}).
These studies are mostly based on the analysis tools and strategies used for the analysis of the partial LHC Run 2 dataset (2015-2016). They do not include further improvements, such as new machine-learning based tools or background estimation strategies, implemented for the later analyses of the full LHC Run 2 dataset.
A full review of the results of these sensitivity studies across the different final states and models is beyond the scope of this article but a few general observations can be made.
Overall, both the increase in integrated luminosity (HL-LHC) and centre-of-mass energy (HE-LHC) lead to a significant sensitivity increase across the different final states. For example, the mass range for a (pseudo)scalar mediator expected to be excluded by \met+\ttbar searches in the simplified model of Section~\ref{sec:SPS_model} with $g=g_q=g_{\chi}=1.0$ (compare Figure~\ref{fig:SPS_limits}) is expected to increase by a factor of two for the HL-LHC compared to the expected sensitivity for LHC Run 3, and by another factor of two for the HE-LHC compared to the HL-LHC.

The sensitivity of most of the searches is dominated by the systematic uncertainties on the main (often irreducible) background processes, for example $\ttbar+V$ in the case of \met+\ttbar searches. In \ttbar final states, these typically arise from two sources: firstly, uncertainties related to reconstructed objects, such as the energy scale for hadronic jets, and, secondly, uncertainties arising from the modelling of SM processes, such as missing higher-order corrections. These uncertainties can vary between a few percent and a few tens of percent, depending on the process and kinematic region. The former are expected to decrease with increasing integrated luminosity as the statistical uncertainties on the measurements from which they are derived are reduced accordingly. A further reduction of these uncertainties can be expected due to the development of better and more refined calibration methods.
The latter can be reduced significantly through profiling in a likelihood fit to data if appropriate, background-enriched control regions are defined. Improved theoretical predictions, for example for differential cross-sections at higher orders in perturbation theory, can also significantly boost the sensitivity of many searches.

In the case of the HE-LHC, in addition to the improvements due to the larger integrated luminosity, the larger centre-of-mass energy provides access to mediator masses beyond the kinematic reach of the (HL-)LHC and to process with small signal cross-sections.

\subsection{FCC-hh}
\label{sec:out_FCChh}

Similar considerations as for the HE-LHC apply to the case of a potential future hadron collider operating at centre-of-mass energies beyond that of the LHC and HE-LHC. The most prominent example is that of the FCC-hh, the Future Circular Collider in its operation mode as a hadron collider with a centre-of-mass energy of $\sqrt{s}=100$~TeV~\cite{FCChh}.
Few dedicated studies regarding the sensitivity of DM searches with top quarks at the FCC-hh exist. For example, in Ref.~\cite{Dutta:2017sod} the sensitivity of the 2-lepton \met+\ttbar final state to Higgs portal models and their extensions is discussed. In general, a significant increase in the accessible mass range of both mediators and DM particles is expected, as well as a significant increase in the sensitivity to smaller DM-SM couplings, rendering detector signatures involving decays of long-lived particles away from the interaction point highly relevant. Moreover, top quarks appearing in the final states of FCC-hh collision can be extremely boosted, underlining the need for high-resolution detectors to identify very collimated decays, as well as the use of advanced pattern recognition methods for top-quark tagging.
A particularly interesting observation is the fact that associated production of a single Higgs boson with \ttbar becomes the dominant Higgs boson production mode at Higgs boson transverse momenta of 1-2~TeV and above, a kinematic regime that would be well-populated at the FCC-hh~\cite{PHarris_2017}. According to initial studies~\cite{PHarris_2017}, searches for invisible Higgs boson decays in this production mode would feature a very low background contamination ($S/B \sim 1$) and hence provide excellent sensitivity to Higgs portal models with small couplings. The corresponding final state would be \met+\ttbar with highly boosted top quarks.

\subsection{Future $e^+e^-$ colliders}
\label{sec:out_FCCee}

No studies of DM searches with top quarks exist for future $e^+e^-$ colliders, such as the International Linear Collider (ILC)~\cite{Behnke:2013xla}, the Compact Linear Collider (CLIC)~\cite{CLIC1,CLIC2}, the Future Circular Collider FCC-ee~\cite{FCCeePhys,FCCee}, and the Circular Electron-Positron Collider (CEPC)~\cite{CEPCStudyGroup:2018rmc,CEPCStudyGroup:2018ghi}. This can be mostly attributed to the fact that these machines are primarily designed for Higgs boson and top quark precision measurements rather than a broad range of BSM (including DM) searches and that their foreseen centre-of-mass energies are in many cases below or close to the \ttbar production threshold.
For example, operation modes at $\sqrt{s}=240$~GeV (250~GeV), i.e. around the maximum of the $Zh$ production cross-section, are foreseen for the FCC-ee and the CEPC (ILC). Additional operation modes in the range 350-365~GeV (FCC-ee, CEPC) and 380~GeV (CLIC) are foreseen for top quark precision measurements. Higher centre-of-mass energies of 1 TeV (ILC) and 1-3~TeV could be possible for the linear $e^+e^-$ machines to allow for wider range of BSM searches.
Hence direct DM production in association with at least one top quark, leading to \met+\ttbar and \met+$t$+$X$ final states, while in principle possible, is trivially limited by the available centre-of-mass energy. Nevertheless, the foreseen precision scans of the \ttbar production threshold at the FCC-ee could in principle be sensitive to anomalous resonant or non-resonant \ttbar production linked with DM or DM mediators as well as anomalous top-quark decays.
Further studies are needed to understand the prospects for DM searches with top quarks at future $e^+e^-$ colliders.

\subsection{Conclusion}
\label{sec:Conclusion}

Collider signatures with top quarks provide sensitive probes of DM predicted by a wide range of models, and possibly even to DE signatures. Searches targeting top-quark production in association with DM or via visible decays of mediator particles have been performed by the ATLAS and CMS Collaborations, with many searches on the full LHC Run 2 collision data still on-going.
As shown in this review, DM searches involving top quarks often provide sensitivity in parameter regions not covered by other DM searches, underlining their importance as sensitive probes of DM at colliders. The upcoming LHC Run 3 opens up further opportunities to improve upon existing results or to explore new signatures, for example involving top quarks in association with long-lived particle signatures.

\section*{Acknowledgements}
K.B.thanks the Helmholtz Association for the support through the ”Young Investigator Group” initiative. The authors acknowledge funding by the Deutsche Forschungsgemeinschaft (DFG, German Research Foundation) under Germany‘s Excellence Strategy – EXC 2121 “Quantum Universe” – 390833306.

\bibliographystyle{unsrtnat}


\begin{thebibliography}{1}

\bibitem{lensingone}
V. Trimble, 
Annu. Rev. Astron. Astrophys. {\bf 25} (1987) 425. 

\bibitem{lensingtwo}
G. Bertone, {\it et al.} 
Phys. Rep. {\bf 405} (2005) 279. 

\bibitem{lensingthree}
J. L. Feng, 
Annual Review of Astronomy and Astrophysics {\bf 48} (2010) 495. 

\bibitem{cmbone}
G. Hinshaw {\it et al.},
Astrophys. J. Suppl. {\bf 208} (2013) 19.

\bibitem{cmbtwo}
Planck Collaboration, 
Astron. Astrophys. {\bf 641} (2020) A1. 

\bibitem{wimp}
G. Steigman {\it et al.},
Nucl. Phys. B {\bf 253} (1985) 375. 

\bibitem{LHC}
L. Evans {\it et al.}, 
JINST {\bf 3} (2008) S08001. 

\bibitem{atlas}
ATLAS Collaboration, 
JINST \textbf{3} (2008) S08003.

\bibitem{cms}
CMS Collaboration, 
JINST \textbf{3} (2008) S08004.

\bibitem{topcdf}
CDF Collaboration, 
Phys. Rev. Lett. \textbf{74} (1995) 2626.

\bibitem{topd0}
D0 Collaboration, 
Phys.~Rev.~Lett.~\textbf{74} (1995) 2632.

\bibitem{SupernovaSearchTeam:1998fmf}
A.~G.~Riess \textit{et al.} [Supernova Search Team],
Astron. J. \textbf{116} (1998) 1009.

\bibitem{SupernovaCosmologyProject:1998vns}
S.~Perlmutter \textit{et al.} [Supernova Cosmology Project],
Astrophys. J. \textbf{517} (1999) 565.

\bibitem{Abercrombie:2015wmb}
{ATLAS/CMS Dark Matter Forum},
Phys. Dark Univ. \textbf{27} (2020) 100371.

\bibitem{ATLAS:2019wdu}
ATLAS Collaboration,
JHEP \textbf{05} (2019) 142.

\bibitem{Kamenik:2011nb}
{J.~F.~Kamenik \it{et al.}},
Phys.~Rev.~D~\textbf{84} (2011) 111502.

\bibitem{Boucheneb:2014wza}
{I.~Boucheneb \it{et al.}},
JHEP \textbf{01} (2015) 017.

\bibitem{ATLAS:2018alq}
{ATLAS Collaboration},
JHEP \textbf{12} (2018) 039.

\bibitem{DAmbrosio:2002vsn}
G.~D'Ambrosio \textit{et al.},
Nucl. Phys. B \textbf{645} (2002) 155. 

\bibitem{Buckley:2014fba}
M.~R.~Buckley \textit{et al.}
Phys. Rev. D \textbf{91} (2015) 015017.

\bibitem{MSSM1}
P. Fayet,
Phys. Lett. B \textbf{64} (1976) 159.

\bibitem{MSSM2}
P. Fayet,
Phys. Lett. B \textbf{69} (1977) 489

\bibitem{Djouadi:2008gy}
A.~Djouadi,
Eur. Phys. J. C \textbf{59} (2009) 389.

\bibitem{Carena:2015uoe}
{M.~Carena \it{et al.}},
JHEP \textbf{02} (2016) 123.

\bibitem{Fuchs:2017wkq}
{E.~Fuchs \it{et al.}},
Eur. Phys. J. C \textbf{78} (2018) 87.

\bibitem{PDG2020}
{Particle Data Group},
Prog. Theor. Exp. Phys. \textbf{2020} (2020) 083C01.

\bibitem{Branco:2011iw}
G.~C.~Branco \textit{et al.},
Phys. Rept. \textbf{516} (2012) 1.

\bibitem{Gunion:2002zf}
J.~F.~Gunion \textit{et al.},
Phys. Rev. D \textbf{67} (2003) 075019.

\bibitem{Berlin:2014cfa}
A.~Berlin \textit{et al.},
JHEP \textbf{06} (2014) 078.

\bibitem{Bell:2016ekl}
N.~F.~Bell \textit{et al.},
JCAP \textbf{03} (2017) 015.

\bibitem{Bauer:2017ota}
M.~Bauer \textit{et al.}
JHEP \textbf{05} (2017) 138.

\bibitem{Goncalves:2016iyg}
D.~Goncalves \textit{et al.}
Phys. Rev. D \textbf{95} (2017) 055027.

\bibitem{Abe:2018emu}
T.~Abe \textit{et al.}
JHEP \textbf{02} (2019) 028.

\bibitem{LHCDarkMatterWorkingGroup:2018ufk}
{LHC Dark Matter Working Group},
Phys. Dark Univ. \textbf{27} (2020) 100351.

\bibitem{ATLAS:2HDMa_2021}
{ATLAS Collaboration},
ATLAS-CONF-2021-036,
\url{https://atlas.web.cern.ch/Atlas/GROUPS/PHYSICS/CONFNOTES/ATLAS-CONF-2021-036}.

\bibitem{Horndeski:1974}
G.W.~Horndeski,
Int. J. Theor. Phys. \textbf{10} (1974) 363. 

\bibitem{Brax:2016did}
P.~Brax \textit{et al.},
Phys. Rev. D \textbf{94} (2016) 084054.

\bibitem{Joyce:2014kja}
A.~Joyce \textit{et al.},
Phys. Rept. \textbf{568} (2015) 1.


\bibitem{ATLAS-CONF-2022-036}
{ATLAS Collaboration},
ATLAS-CONF-2022-036,
\url{https://atlas.web.cern.ch/Atlas/GROUPS/PHYSICS/CONFNOTES/ATLAS-CONF-2022-036/}

\bibitem{CMS:2018gbj} 
{CMS Collaboration},
JHEP \textbf{06} (2018) 027.

\bibitem{ATLAS:2018wis}
{ATLAS Collaboration},
Eur. Phys. J. C \textbf{79} (2019) 375.

\bibitem{ATLAS:DNNTopPerf}
{ATLAS Collaboration},
ATL-PHYS-PUB-2020-017,
\url{https://atlas.web.cern.ch/Atlas/GROUPS/PHYSICS/PUBNOTES/ATL-PHYS-PUB-2020-017/}.

\bibitem{XGBoost}
{T. Chen \it{et al.}},
KDD \textbf{16} (2016) 785.

\bibitem{Cacciari:2008gp}
{M. Cacciari \it{et al.}}, 
JHEP \textbf{04} (2008) 063.

\bibitem{Thaler:2010tr}
{J.~Thaler \it{et al.}},
JHEP \textbf{03} (2011) 015. 

\bibitem{Larkoski:2013eya}
{A.~J.~Larkoski \it{et al.}},
JHEP \textbf{06} (2013) 108.

\bibitem{Moult:2016cvt}
{I.~Moult \it{et al.}},
JHEP \textbf{12} (2016) 153. 

\bibitem{Hocker:2007ht}
A.~Hoecker \textit{et al.}, [arXiv:0703039[data-an]]. 

\bibitem{ATLAS:2018cjd}
{ATLAS Collaboration},
JHEP \textbf{05} (2019) 041.

\bibitem{CMS:2019zzl}
{CMS Collaboration},
JHEP \textbf{03} (2019) 141.

\bibitem{ATLAS:2020yzc}
{ATLAS Collaboration},
Eur. Phys. J. C \textbf{81} (2021) 860.

\bibitem{ATLAS:METplusTopW}
{ATLAS Collaboration},
ATLAS-CONF-2022-012
\url{https://atlas.web.cern.ch/Atlas/GROUPS/PHYSICS/CONFNOTES/ATLAS-CONF-2022-012/}

\bibitem{CMS:2021eha}
CMS Collaboration,
Eur.~Phys.~J.~C {\bf 81} (2021) 970.

\bibitem{cms_stop_0l}
CMS Collaboration, 
Phys. Rev. D {\bf 104} (2021) 052001.
\bibitem{cms_stop_1l}
CMS Collaboration, 
JHEP {\bf 05} (2020) 032.

\bibitem{cms_stop_2l} 
CMS Collaboration, 
Eur. Phys. J. C {\bf 81} (2021) 3. 

\bibitem{cms_deepak8}
CMS Collaboration, 
JINST {\bf 15} (2020) P06005. 

\bibitem{cms_metsig}
CMS Collaboration, 
JINST {\bf 14} (2019) P07004. 

\bibitem{cms_stransversemass}
CMS Collaboration, 
Phys. Rev. D {\bf 97} (2018) 032009. 

\bibitem{ATLAS:2020dsf}
{ATLAS Collaboration},
Eur. Phys. J. C \textbf{80} (2020) 737.

\bibitem{ATLAS:2020xzu}
{ATLAS Collaboration},
JHEP \textbf{04} (2021) 174.

\bibitem{ATLAS:2021hza}
{ATLAS Collaboration},
JHEP \textbf{04} (2021) 165.

\bibitem{ATLAS:METplustt}
 {ATLAS Collaboration},
[arXiv:2211.05426 [hep-ex]].




\bibitem{cms_exo_18_010}
CMS Collaboration,
JHEP {\bf 03} (2019) 141.

\bibitem{ATLAS:2020lks}
{ATLAS Collaboration},
JHEP \textbf{10} (2020) 061.

\bibitem{ATLAS:2018rvc}
{ATLAS Collaboration},
Eur. Phys. J. C \textbf{78} (2018) 565.

\bibitem{CMS:2018rkg}
{CMS Collaboration},
JHEP \textbf{04} (2019) 031.

\bibitem{ATLAS:2017snw}
{ATLAS Collaboration},
Phys. Rev. Lett. \textbf{119} (2017) 191803.

\bibitem{CMS:2019pzc}
{CMS Collaboration},
JHEP \textbf{04} (2020) 171.

\bibitem{ATLAS:2020hpj}
{ATLAS Collaboration},
Eur. Phys. J. C \textbf{80} (2020) 1085.

\bibitem{ATLAS:2022rws}
{ATLAS Collaboration},
[arXiv:2211.01136 [hep-ex]].

\bibitem{Baldi:2016fzo}
P.~Baldi \textit{et al.},
Eur. Phys. J. C \textbf{76} (2016) 235.

\bibitem{ATLAS:2017oes}
{ATLAS Collaboration},
JHEP \textbf{09} (2017) 088.

\bibitem{ATLAS:ttAH}
{ATLAS Collaboration},
ATLAS-CONF-2022-008,
\url{https://atlas.web.cern.ch/Atlas/GROUPS/PHYSICS/CONFNOTES/ATLAS-CONF-2022-008/}.

\bibitem{CMS:2019rvj}
{CMS Collaboration},
Eur. Phys. J. C \textbf{80} (2020) 75.

\bibitem{ATLAS:2021upq}
{ATLAS Collaboration},
JHEP \textbf{06} (2021) 145.

\bibitem{CMS:2020imj}
{CMS Collaboration},
JHEP \textbf{07} (2020) 126.

\bibitem{CMS:2019rlz}
{CMS Collaboration},
JHEP \textbf{01} (2020) 096.

\bibitem{ATLAS:ttZp}
{ATLAS Collaboration},
ATLAS-CONF-2021-048,
\url{https://atlas.web.cern.ch/Atlas/GROUPS/PHYSICS/CONFNOTES/ATLAS-CONF-2021-048/}.

\bibitem{CMS:DMSum}
{CMS Collaboration},
\url{https://twiki.cern.ch/twiki/bin/view/CMSPublic/SummaryPlotsEXO13TeV#DM_summary_plots}.

\bibitem{ATLAS:DMSum}
{ATLAS Collaboration},
ATL-PHYS-PUB-2022-036,
\url{https://atlas.web.cern.ch/Atlas/GROUPS/PHYSICS/PUBNOTES/ATL-PHYS-PUB-2022-036/}.

\bibitem{CMS:2021far}
CMS Collaboration, 
JHEP \textbf{11} (2021) 153.

\bibitem{CMS:2019emo}
CMS Collaboration,
Phys. Rev. D \textbf{100} (2019) 11. 

\bibitem{CMS:2019gwf}
CMS Collaboration,
JHEP \textbf{05} (2020) 033.

\bibitem{Haisch:2018bby}
{U.~Haisch \it{et al.}},
JHEP \textbf{02} (2019) 029.

\bibitem{Pani:2017qyd}
{P.~Pani \it{et al.}},
Phys. Dark Univ. \textbf{21} (2018) 8.

\bibitem{Pinna:2017tay}
{D.~Pinna \it{et al.}},
Phys. Rev. D \textbf{96} (2017) 035031.



\bibitem{ALPs1}
{M.~B.~Gavela \it{et al.}},
Phys. Lett. B \textbf{169} (1986) 73.

\bibitem{ALPs2}
{K. Choi \it{et al.}},
Phys. Lett. B \textbf{181} (1986) 145.

\bibitem{Gavela:2019cmq}
{M.~B.~Gavela \it{et al.}},
Phys. Rev. Lett. \textbf{124} (2020) 051802.

\bibitem{CMS:2021xor}
{CMS Collaboration}
JHEP \textbf{04} (2022) 087.

\bibitem{Carra:2021ycg}
{S.~Carra \it{et al.}},
Phys. Rev. D \textbf{104} (2021) 092005.

\bibitem{Brivio:2017ije}
{I.~Brivio \it{et al.}},
Eur. Phys. J. C \textbf{77} (2017) 572.

\bibitem{Carmona:2022jid}
{A.~Carmona \it{et al.}},
JHEP \textbf{07} (2022) 122.

\bibitem{Renner:2018fhh}
{S.~Renner \it{et al.}},
JHEP \textbf{08} (2018) 052.

\bibitem{FVALPs}
{C. Froggatt \it{et al.}},
Nucl. Phys. B \textbf{147} (1979) 277.

\bibitem{ALPS_NR_Feyn}
{J. Bonilla \it{et al.}},
\url{https://ep-news.web.cern.ch/content/non-resonant-searches-open-new-avenues-hunt-axion-particles}.

\bibitem{Kribs:2018ilo}
{G.~D.~Kribs \it{et al.}},
JHEP \textbf{07} (2019) 133.

\bibitem{Haisch:2021ugv}
{U.~Haisch \it{et al.}},
JHEP \textbf{09} (2021) 206.

\bibitem{HLLHC_YR}
{A.~Dainese \it{et al.}},
CERN-2019-007,
\url{https://cds.cern.ch/record/2703572}.

\bibitem{FCChh}
{M. Benedikt \it{et al.}},
CERN-ACC-2018-0058,
\url{https://cds.cern.ch/record/2651300}.

\bibitem{Dutta:2017sod}
{B.~Dutta \it{et al.}},
Eur. Phys. J. C \textbf{78} (2018) 595.

\bibitem{PHarris_2017}
{P.~Harris \it{et al.}},
Talk at FCC Week 2017,
\url{https://indico.cern.ch/event/556692/contributions/2592531/attachments/1469109/2272523/PCH_FCC_DM_v2.pdf}.

\bibitem{Behnke:2013xla}
{T.~Behnke \it{et al.}},
[arXiv:1306.6327[acc-ph]].

\bibitem{CLIC1}
{M.~Aicheler \it{et al.}},
CERN Yellow Reports: Monographs 2012,
\url{https://cds.cern.ch/record/1500095}.

\bibitem{CLIC2}
{L.~Linssen \it{et al.}},
[arXiv:1202.5940[ins-det]].

\bibitem{FCCeePhys}
{A.~Abada \it{et al.}},
Eur. Phys. J. C \textbf{79} (2019) 474. 

\bibitem{FCCee}
{A.~Abada, \it{et al.}},
Eur. Phys. J. S. T. \textbf{228} (2019) 261. 

\bibitem{CEPCStudyGroup:2018rmc}
{CEPC Study Group},
[arXiv:1809.00285[acc-ph]].

\bibitem{CEPCStudyGroup:2018ghi}
{CEPC Study Group},
[arXiv:1811.10545[hep-ex]].




\end{thebibliography}

\end{document}